\documentclass[journal]{new-aiaa}
\usepackage[utf8]{inputenc}
\usepackage{textcomp}
\usepackage{orcidlink}
\usepackage{graphicx}
\usepackage{amsmath}
\usepackage[version=4]{mhchem}
\usepackage{siunitx}

\graphicspath{{/}}
\usepackage{booktabs}
\usepackage{braket}
\usepackage{bm}
\usepackage{xspace}
\usepackage{subcaption}
\usepackage{tikz}
\usepackage{ulem}
\usepackage{pgfplots}
\usepgfplotslibrary{colorbrewer,external,fillbetween}
\pgfplotsset{
    compat=newest,
    width=0.75\textwidth,
    height=0.45\textwidth,
    scale only axis,
    xlabel=$x$,
    ylabel=$y$,
    yticklabel style={text width=2em},
    grid=major,
    grid style={line width=.1pt, draw=gray!50},
	cycle list/Set1-9,
	cycle multiindex* list={
		Set1-9\nextlist
		mark list*\nextlist
		solid, every maker/.append style={solid, draw=.!80!black, fill=.}\nextlist
	},
    every axis/.append style={line width = 1.3pt},
     colormap={inferno}{
     rgb(0)=(0.001462, 0.000466, 0.013866),
     rgb(15)=(0.037668, 0.025921, 0.132232),
     rgb(30)=(0.116656, 0.047574, 0.272321),
     rgb(45)=(0.217949, 0.036615, 0.383522),
     rgb(60)=(0.316282, 0.053490, 0.425116),
     rgb(75)=(0.410113, 0.087896, 0.433098),
     rgb(90)=(0.503493, 0.121575, 0.423356),
     rgb(105)=(0.596940, 0.154848, 0.398125),
     rgb(120)=(0.688653, 0.192239, 0.357603),
     rgb(135)=(0.775059, 0.239667, 0.303526),
     rgb(150)=(0.851384, 0.302260, 0.239636),
     rgb(165)=(0.912966, 0.381636, 0.169755),
     rgb(180)=(0.956852, 0.475356, 0.094695),
     rgb(195)=(0.981895, 0.579392, 0.026250),
     rgb(210)=(0.987464, 0.690366, 0.079990),
     rgb(225)=(0.973088, 0.805409, 0.216877),
     rgb(240)=(0.947594, 0.917399, 0.410665),
     rgb(255)=(0.988362, 0.998364, 0.644924),
     },
}
\usetikzlibrary{quantikz2,shapes,external}
\newcommand{\Fig}[1]{\mbox{Fig.~\ref{#1}}} 
\newcommand{\Eq}[1]{\mbox{Eqn.~\eqref{#1}}}
\renewcommand{\O}{\Omega} 
\newcommand{\eb}{\mbox{$\bm{e}$}\xspace} 
\newcommand{\upb}{\mbox{$\bm{\upsilon}$}\xspace} 
\newcommand{\lamb}{\mbox{$\bm{\lambda}$}\xspace}
\newcommand{\ub}{\mbox{$\bm{u}$}\xspace} 
\newcommand{\Cb}{\mbox{$\bm{C}$}\xspace} 
\newcommand{\Ab}{\mbox{$\bm{A}$}\xspace} 
\newcommand{\pb}{\mbox{$\bm{p}$}\xspace} 
\newcommand{\mb}{\mbox{$\bm{m}$}\xspace} 
\newcommand{\yb}{\mbox{$\bm{y}$}\xspace} 

\definecolor{red1}{RGB}{0,0,0}
\definecolor{red2}{RGB}{0,0,0}
\definecolor{blue}{RGB}{0,0,0}
\definecolor{green}{RGB}{0,0,0}
\definecolor{purple}{RGB}{0,0,0}
\definecolor{orange}{RGB}{0,0,0}
\definecolor{yellow}{RGB}{0,0,0}
\definecolor{brown}{RGB}{0,0,0}
\definecolor{pink}{RGB}{0,0,0}
\definecolor{gray}{RGB}{170,170,170}
\definecolor{lightgray}{RGB}{230,230,230}
\definecolor{darkgray}{RGB}{120,120,120}

\title{Towards Variational Quantum Algorithms for generalized linear and nonlinear transport phenomena}

\author{Sergio Bengoechea 
\orcidlink{0009-0001-8205-5878} 
\footnote{Research Associate, Institute for Fluid Dynamics and Ship Theory, Hamburg University of Technology, Am Schwarzenberg-Campus 4, D-21073 Hamburg, Germany; \href{mailto:sergio.bengoechea@tuhh.de}{sergio.bengoechea@tuhh.de}. (Corresponding author)} and 
Paul Over
\orcidlink{0000-0001-7436-5254}
\footnote{Research Associate, Institute for Fluid Dynamics and Ship Theory, Hamburg University of Technology, Am Schwarzenberg-Campus 4, D-21073 Hamburg, Germany; \href{mailto:paul.over@tuhh.de}{paul.over@tuhh.de}. (Co-Corresponding author)}}
\affil{Hamburg University of Technology, D-21703 Hamburg, Germany}

\author{Dieter Jaksch
\orcidlink{0000-0002-9704-3941}
\footnote{Professor, Institute for Quantum Physics, University of Hamburg, Luruper Chaussee 149, D-22761 Hamburg, Germany; \href{mailto:dieter.jaksch@uni-hamburg.de}{dieter.jaksch@uni-hamburg.de}.}}
\affil{University of Hamburg, 22761 Hamburg, Germany}
\affil{University of Oxford, Oxford OX1 3PU, United Kingdom}

\author{Thomas Rung
\orcidlink{0000-0002-3454-1804}
\footnote{Professor, Institute for Fluid Dynamics and Ship Theory, Hamburg University of Technology, Am Schwarzenberg-Campus 4, D-21073 Hamburg, Germany; \href{mailto:thomas.rung@tuhh.de}{thomas.rung@tuhh.de}.}}
\affil{Hamburg University of Technology, D-21703 Hamburg, Germany}

\begin{document}

\maketitle

% ABSTRACT
\begin{abstract}
This article proposes a Variational Quantum Algorithm to solve linear and nonlinear thermofluid dynamic transport equations. The hybrid classical-quantum framework is applied to problems governed by the heat, wave, and Burgers' equation in combination with different engineering boundary conditions. Topics covered include the encoding of band matrices, as in the consideration of non-constant material properties and upwind-biased first- and higher-order approximations, widely used in engineering Computational Fluid Dynamics, by the use of a mask function. Verification examples demonstrate high predictive agreement with classical methods. Furthermore, the scalability analysis shows a \textit{polylog} scaling of the number of quantum gates with the number of qubits. Remaining challenges refer to the implicit construction of upwind schemes and the identification of an appropriate parameterization strategy of the quantum ansatz.
\end{abstract}
% ABSTRACT

%###########################################
%       Intro
%###########################################
\section{Introduction}
\label{sec:Intro}
Computational Fluid Dynamics (\textsc{CFD}) has been established as a central discipline in computational science and engineering with applications ranging from climate research to energy conversion or transportation and biomedical industries. The interest in numerical resolution of larger spatial and temporal scales encounters extremely expensive and energy-intensive simulations on classical hardware \cite{Nasa2030}. In this respect, Quantum Computers (\textsc{QC}s) promise a computational potential that cannot be achieved by classical hardware \cite{Jaksch2023a}. 

The research field of Quantum \textsc{CFD} (\textsc{QCFD}) is rapidly growing and can be divided into quantum-inspired approaches~\cite{Gourianov2022,Kiffner2023,Peddinti2024,Gourianov2024}, which are not necessarily implemented on quantum hardware, and methods that are closely oriented to the capabilities and restrictions of quantum hardware \cite{Steijl2018,Cao2013, Harrow2009,Gaitan2020,Oz2021,Kacewicz2006,Chen2021,Brearley2024,Over2024b,Sato2024,Hu2024,Wright2024,Esmaeilifar2024,Suau2021,Preskill2018,Kyriienko2021,Demirdjian2022,BravoPrieto2023,Sato2021,Leong2022,Leong2023,Leong2024,Ingelmann2024,Over2024}. The present work focuses on the second type of methods. Excellent recent summaries of the advantages and disadvantages of different \textsc{QCFD} methods and their predictive prospects can be found in Refs.~\cite{Bharadwaj2020,Bharadwaj2023}. We therefore limit ourselves to a brief summary of prior efforts below.

\citeauthor{Steijl2018}~\cite{Steijl2018} used a hybrid classical-quantum approach for solving the Poisson problem in the vortex-in-cell method via a quantum Fourier algorithm. A similar strategy is employed in the work of \citeauthor{Cao2013}~\cite{Cao2013} to approximate the Poisson equation in the predictor-corrector solver of the incompressible Navier-Stokes equation. Here, the Quantum Linear Solver of Equations (\textsc{QLSE}) proposed by \citeauthor*{Harrow2009}~(\textsc{HHL})~\cite{Harrow2009} is applied. In Ref.~\cite{Gaitan2020} and \cite{Oz2021}, the Quantum Amplitude Estimation Algorithm (\textsc{QAEA}), described by \citeauthor{Kacewicz2006}~\cite{Kacewicz2006}, solves the system of coupled Ordinary Differential Equations (\textsc{ODE}s), following the spatial discretization of the Navier-Stokes and the Burgers' equation, respectively. The incremental time application of the \textsc{HHL} method is employed by \citeauthor{Chen2021}~\cite{Chen2021} to approximate compressible flows after discretizing the Navier-Stokes equation with classical finite volumes. Another methodology employs the Hamiltonian simulation for transport problems as it has been proposed by \citeauthor{Brearley2024}~\cite{Brearley2024} for the advection equation and \citeauthor{Over2024b}~\cite{Over2024b} for the advection-diffusion equation. Instead of the \textsc{QLSE} approach, the system dynamics of the discrete problem are herein unitarized for a time evolution strategy, which is suitable for \textsc{QC}s. 
In this regard, the efficiency of the simulation can be improved using a Bell basis, facilitating the diagonalization of Hamiltonians as introduced by \citeauthor{Sato2024}~\cite{Sato2024}. The approach employs an operator splitting method, which is also used by \citeauthor{Hu2024}~\cite{Hu2024} to implement a spectral method that handles Dirichlet and periodic boundary conditions. In particular, when applying a spectral method to simulate the wave equation with periodic boundary conditions, a favorable single-diagonal operator naturally arises, as presented by \citeauthor{Wright2024}~\cite{Wright2024}. 
For modeling nonlinearities, for example, those from the Burgers' equation, \citeauthor{Esmaeilifar2024}~\cite{Esmaeilifar2024} present a fractal structure of multiple state preparations that grows exponentially with the number of time steps. Unfortunately, such algorithms have to devote a range of qubits to error corrections, which significantly increases the total qubit count. Therefore, these algorithms are more appropriate for future fault-tolerant \textsc{QC} than for the current Noise Intermediate-Scale Quantum (\textsc{NISQ}) era \cite{Suau2021,Preskill2018}. 

Variational Quantum Algorithms (\textsc{VQA}s) provide an alternative for solving \textsc{CFD} problems on \textsc{NISQ} devices. \textsc{VQA} procedures rely on the combination of classical and quantum hardware, and require reformulating the governing Partial Differential Equation (\textsc{PDE}) as an optimization problem. The \textsc{QC} part of the employed hardware is dedicated to efficiently evaluating the cost function, while the classical hardware drives the optimizer for a parameterized solution. 
The hybrid quantum-classical characteristic highlights that \textsc{QC}s do not need to replace classical computing entirely to provide a computational advantage. 
The main advantage of \textsc{VQA}s is the use of shallow circuits with a small number of gate operations, thereby also reducing potential decoherence errors. By comparing highly optimized classical \textsc{CFD} simulations with unoptimized \textsc{VQA} prototypes, \citeauthor{Syamlal2024}~\cite{Syamlal2024} estimate quantum advantage for discretizations beyond $10^8$ grid points. This projection suggests that \textsc{VQA} methods could become increasingly viable, particularly for large-scale \textsc{CFD} applications.

Recent applications of \textsc{VQA}s for fluid dynamics are reported in Refs.~ \cite{Demirdjian2022,BravoPrieto2023,Kyriienko2021,Siegl2025b,Sato2021,Leong2022,Leong2023,Leong2024,Ingelmann2024,Over2024,Bharadwaj2020,Bharadwaj2023}. Some of these approaches are inspired by Physics Informed Neural Networks (\textsc{PINN}s) \cite{Raissi2019}, where quantum neural networks are trained variationally \cite{Kyriienko2021,Siegl2025b} to solve the Navier-Stokes equation for supersonic nozzle flows. In contrast, the work by \citeauthor{Demirdjian2022}~\cite{Demirdjian2022} separates the discretized advection-diffusion equation into a finite series of unitary operations that are approximated with the variational quantum linear solver introduced in Ref.~\cite{BravoPrieto2023}. \citeauthor{Sato2021}~\cite{Sato2021} presents a potential energy minimization of the elliptic Poisson equation using a \textsc{VQA} that linearly decomposes the equation's dynamics in a series of parameterized quantum circuits. This energy minimization approach is adapted to time-dependent problems in Refs.~\cite{Leong2022,Leong2023,Leong2024}. Alternatively, \citeauthor{Ingelmann2024}~\cite{Ingelmann2024} uses a least-square formulation of the cost function, doubling the number of terms in the optimization. To reduce the effort, \citeauthor{Over2024}~\cite{Over2024} propose an energy minimization approach and extend the Hadamard test-based \textsc{VQA}, described in \citeauthor{Lubasch2020}~\cite{Lubasch2020}, to the treatment of \textsc{PDE}s in combination with a variety of engineering boundary conditions. 

The present investigation extends the scope of Ref.~\cite{Over2024} to more general convection-diffusion-reaction problems in combination with engineering boundary conditions. 
Accordingly, we propose a hybrid \textsc{VQA} based on Refs.~\cite{Lubasch2020,Over2024} for current \textsc{NISQ} architectures, where the \textsc{QC} accelerates the most demanding computational parts.
The framework is able to convert arbitrary band matrices resulting from the discretization of \textsc{PDE}s on structured grids into native quantum gates with \textit{polylog} gate complexity in the number of qubits. 
To this end, different approximations of convective kinematics frequently used in engineering \textsc{CFD}, such as Central-Differencing-Scheme (\textsc{CDS}), Upwind-Differencing-Scheme (\textsc{UDS}) or the higher-order Linear-Upwind-Differencing-Scheme (\textsc{LUDS}), and Quadratic-Upstream-Interpolation-for-Convective-Kinematics (\textsc{QUICK}), are integrated into the Hadamard test-based \textsc{VQA} framework, additionally, the occurrence of inhomogeneous material properties is also considered. 

The publication is structured as follows: Section~\ref{sec:Math} describes the mathematical model and outlines the reformulation of the \textsc{PDE} into a minimization problem. The general layout of the \textsc{VQA} and its major building blocks are explained in Sec.~\ref{sec:Quantum}, while Sec.~\ref{sec:QCFD} provides details of the quantum circuits representing the upwind-biased convective kinematics as well as the inhomogeneous material properties. An in-depth complexity analysis is given in Sec.~\ref{sec:qcfd:gatecomplexity}. The \textsc{VQA} results obtained for the wave, heat, and Burgers' equations are compared with classical approaches in Sec.~\ref{sec:Results} and Sec.~\ref{sec:Conclusion} is devoted to conclusions. The quantum circuits presented in this study have been emulated with \textsc{IBM}'s \textsc{Qiskit} environment v.0.42.1~\cite{Qiskit}, following a \textit{little-endian} convention in the binary representation. Throughout the manuscript, symbolic matrices and vectors are displayed in bold.

%###########################################
%       Math
%###########################################
\section{Mathematical Framework}
\label{sec:Math}
The section contains two major parts, i.e., the derivation of the discretized governing equations (Secs.~\ref{sec:governingeq} and \ref{sec:discretize}), and the preparation of the corresponding optimization problem (Sec.~\ref{sec:spatialopt}). 

\subsection{Governing Equations} \label{sec:governingeq}
The presented approach is restricted to second-order linear and nonlinear \textsc{PDE}s, commonly appearing in \textsc{CFD}. The framework focuses on unsteady problems in one-dimensional spatial domains $\O$ and extensions to higher spatial dimensions are straightforward.
We employ non-dimensional properties $y = \hat y/Y$, where $Y$ is a dimensional reference property, and use a dimensionless description of space $\hat{x} \in [\hat x_0,\hat x_1]$ and time $\hat{t} \in [0,T]$ with $x = \hat{x}/(\hat x_1- \hat x_0)=\hat{x}/L$ and $t = \hat{t}/T$. 
Material properties are given by, the constant density $\hat{\rho} \, [\si{\kg\per\cubic\m}{}]$, the dynamic
viscosity $\hat{\mu} \, [\si{\kg\per\m\per\s}{}]$, the speed of sound $\hat{c} \, [\si{\m\per\s}{}]$ and the thermal diffusivity $\hat{\alpha}(x) \, [\si{\square\m\per\s}{}]$.  The governing equations describe the evolution of the dimensionless flow quantities referred to as the state $y$, i.e., the velocity $v = \hat{v} / V = \hat{v} (T/L)$,  the temperature $\vartheta = \hat{\vartheta}/\vartheta_{\text{ref}}$, and the pressure $p = \hat{p}/p_{\text{ref}}$. 

A set of five non-dimensional similarity parameters $a_i(x,t,y)$ that depend on space, time and state, are introduced together with a non-dimensional source term~$f$, to formulate the dimensionless second-order \textsc{PDE}, viz.  
\begin{equation}
    \begin{split}
       a_1 \frac{\partial^2 y}{\partial t^2} +  a_2 \frac{\partial y }{ \partial t} - \frac{\partial}{\partial x} \left( a_3  \frac{\partial y}{\partial x}\right) + a_4  \frac{\partial y}{\partial x } + a_5  y& =  f  \qquad \text{ in } \Omega_\text{T}:=(0,1) \times (0,T]\,,\\
       y(x,t) &= y_\text{BC}   \quad  \text{ on } \partial \Omega\,,\\
       y(x,t) &= y_0  \; \, \;  \quad \text{ at } t = 0\,.
       \label{eq:GoverningEq}
       \end{split}
\end{equation}

The coefficients $a_i$ depend on the governing equation, cf.~Tab.~\ref{tab:Math:Coefficients}. For the example of the non-dimensional Burgers' equation, the state refers to $y = v$ and the coefficients are $a_1 =0,\, a_2=1,\, a_3=1/Re,\, a_4=v\; \text{and } a_5=0$, where  $Re=\hat{\rho}{V}L/\hat{\mu}$ is the \textit{Reynolds} number. The transient heat conduction problem ($y = \vartheta$) is recovered with $a_1=0,\, a_2=1,\, a_3=\alpha(x),\, a_4=0 \, \text{and } a_5=0$, where the dimensionless thermal diffusivity $\alpha(x) = \hat{\alpha}(x) /(VL)$ varies in space. For modeling the propagation of sound ($y = p$), a hyperbolic setting is recovered by choosing $a_1={Ma}^2, \, a_2=0,\, a_3=1,\, a_4=0 \, \text{and } a_5=0$, where $Ma^2 =  V^2/\hat{c}^2$ is the square of the \textit{Mach} number. 

\begin{table}[htbp]
\centering
\caption{Key coefficients {$\pmb{a_i(x,t,y)}$ for the second-order \textsc{PDE} in \Eq{eq:GoverningEq}.}}
\label{tab:Math:Coefficients}
\begin{tabular}{@{}lccccc@{}}
\toprule
Governing equation & $a_1$ & $a_2$ & $a_3$ & $a_4$ & $a_5$ \\ \midrule
Transient heat conduction      & 0 & 1         & $\alpha(x)$  & 0         & 0    \\
Transient convection diffusion & 0 & 1         & $1/Re$      & $v$  & 0     \\ 
Acoustic wave propagation    & $Ma^2$ & 0         & $1$       & 0         & 0    \\\bottomrule
\end{tabular}
\end{table}

\subsection{Spatio-Temporal Discretization}
\label{sec:discretize}
The governing \Eq{eq:GoverningEq} is solved on the unit space-time interval $\Omega_\text{T}$. The domain is  discretized by $N_\text{p}+2$ equidistantly spaced points $x_k : k \in \left[0,N_\text{p}+1\right]$ and $N_\text{t}+1$ equidistant time instants $t^l :  l \in \left[0,N_\text{t}\right]$. The step size designation is $\Delta x$ and $\Delta t$. Following our previous contribution \cite{Over2024}, we distinguish between interior points, where \Eq{eq:GoverningEq} is evaluated, and boundary (ghost) points ($x_0, x_{N_\text{p}+1}$), which are used to implement boundary conditions. The reader is referred to Ref.~\cite{Over2024} for a detailed description of combinations of Dirichlet or Neumann boundary conditions. 

\subsection{Spatial Optimization Problem} \label{sec:spatialopt}
To formulate the optimization problem, we need a time-discrete initial formulation, cf.~\cite{Over2024}. 
In the present paper, temporal derivatives are approximated by the following  exemplary backward Finite Difference (\textsc{FD}) formulae: 
\begin{equation}
         \frac{\partial y}{\partial t}= \frac{ y^l -y^{l-1} }{\Delta t} + \mathcal{O}(\Delta t) \, ,
         \label{eq:timeDiscr}
          \quad
         \frac{\partial^2 y}{\partial t^2} = \frac{ y^l -2y^{l-1} + y^{l-2} }{\Delta t^2} + \mathcal{O}(\Delta t) \,.
\end{equation}

Introducing \Eq{eq:timeDiscr} into \Eq{eq:GoverningEq}, we obtain the semi-discrete residual form
 \begin{equation}
     \begin{split}
   R(x,t^l) = 
  -\frac{\partial}{\partial x} \left( a_3 \frac{\partial y}{\partial x} \right) + a_4  \frac{\partial y}{\partial x } + 
  y^l \left( a_5 + \frac{a_1}{\Delta t^2} + \frac{a_2}{\Delta t}\right)  - \left[ 
 a_1 \left( \frac{2y^{l-1} -y^{l-2}}{\Delta t^2}\right)  
 + a_2  \frac{y^{l-1}}{\Delta t}
 + f \right]
  = 0 \, .
    \label{eq:ResidualEq}
    \end{split}
\end{equation}
The desired weighted residual form of \Eq{eq:ResidualEq} is obtained by 
\begin{equation}
 \int_{\O}  z(x,t^l) \, R(x,t^l) \,  dx  = 0 \, ,
 \label{equ:weighted_res-2}
\end{equation}
where the choice of the weighting function $z$ offers many options. Using $z=R(x,t^l)$, the optimization problem resembles the \textit{Bubnov-Galerkin} approach which returns a least square-type cost function, cf. ~\cite{Lubasch2020}. 
The disadvantage of this formulation is the larger number of terms in the cost function due to the binomial form and the required modeling of quadratic terms with quantum circuits.
We therefore opt for a \textit{Ritz-Galerkin} approach with $z=y(x,t^l)$, which yields shallower quantum circuits and half the cost function contributions. The related solution to \Eq{equ:weighted_res-2}, i.e.,
   \begin{equation}
    \begin{split}
   &\underbrace{\int_\Omega y^l \Bigg[ 
   \left( a_5 + \frac{a_1}{\Delta t^2} + \frac{a_2}{\Delta t}\right) y^l -\frac{\partial}{\partial x} \left( a_3  \frac{\partial y^l}{\partial x} \right) \Bigg] \,dx}_{B(y^l,y^l)}  \\
   &\hspace{3.5cm}+ \underbrace{\int_\Omega y^l \Bigg[a_4  \frac{\partial y^{l-1}}{\partial x }  - \bigg[ 
 a_1 \left( \frac{2y^{l-1} -y^{l-2}}{\Delta t^2}\right)  
 + a_2 \frac{y^{l-1}}{\Delta t}
 + f \bigg] \Bigg]\,dx}_{F(y^l)}
  = 0\, ,
   \label{eq:WResidualEq}
   \end{split}
\end{equation}
is equivalent to the solution of the variational problem \cite{Grossmann2007,Glowinski2015} characterized by an objective function~$J$ that consists of a bilinear term~$B$ and a linear term~$F$, viz. 
\begin{equation}
    \min\limits_{y^l} J(y^l) \quad  \text{with} \quad J(y^l):= B(y^l,y^l) - 2F(y^l)\,.
    \label{eq:Generic_Opt_Approx}
\end{equation}

The residual form in \Eq{eq:ResidualEq} must be adapted to the requirements of \Eq{eq:Generic_Opt_Approx}, i.e., $B$ needs to be positive semi-definite and symmetric, and numerical schemes are therefore usually limited to symmetric approaches \cite{Sato2021,Lubasch2020}. To avoid this limitation, we are motivated to treat the convective term associated with $a_4$ explicitly and to classify it in~$F$ in \Eq{eq:WResidualEq}. Due to the explicit treatment, the stability of the numerical method constrains the time step size in terms of a \textit{Courant} number criterion, i.e.,
$\Delta t \le \Delta x /v $~\cite{Ferziger2020}.

To obtain a discrete formulation, approximations for both the spatial integrals and the spatial derivatives in \Eq{eq:WResidualEq}, must be introduced. Spatial integrals are approximated using the second-order accurate midpoint rule, cf.~\cite{Over2024}. The approximation of spatial derivatives is linked to the design of the quantum circuits and is presented in Sec.~\ref{sec:Quantum}.

%###########################################
%       Quantum
%###########################################
\section{Quantum Model}
\label{sec:Quantum}
The quantum model describes how \textsc{FD} operations with $y$ can be represented using the principles of quantum computing. Our strategy is based, as mentioned above, on a \textsc{VQA} method, for which a comprehensive introduction w.r.t. \textsc{CFD} can be found in Refs.~\cite{Jaksch2023a,Over2024,Bharadwaj2023}. In Secs.~\ref{sec:encode} and \ref{sec:hybrid}, we first discuss the encoding of the state $y$ in a quantum register and subsequently introduce the hybrid quantum-classical framework for realizing the \textsc{VQA}. Due to the stochastic nature of \textsc{QC}s, the evaluation of individual qubits is generally demanding. To reduce the number of evaluations, i.e., measurements, the strategy pursued here is based on two different Hadamard test circuits, one representing the bilinear term~$B$ and one for the linear term~$F$ in \Eq{eq:Generic_Opt_Approx}. In both cases, the discrete integral is determined over all $N_\text{p}$~points of the inner region at a measuring qubit (ancilla), instead of measuring the individual contributions for each qubit. Finally, the fundamental building blocks for the representation of inhomogeneous material properties and upwind-biased differential operators with Quantum-Nonlinear-Processing-Units (\textsc{QNPU}s) are introduced in Sec.~\ref{sec:qnpus}. 

\subsection{State Encoding} \label{sec:encode}
We encode the state amplitudes in a quantum register comprising $n=\log_2\left(N_\text{p}\right)$ qubits that span an orthonormal basis in the complex Hilbert space by the tensor product of each qubit's subspace. The $i$-th element of the basis is given in the binary representation $\text{bin}(i)$ as $\ket{\text{bin}(i)} = \eb_i$, e.g., for the \textit{little-endian} convention $\eb_{i=0} = \ket{0}\otimes\ket{0}\otimes~\hdots~\otimes~\ket{0} $ or $\eb_{i=1} = \ket{0}\otimes\ket{0}\otimes~\hdots~\otimes~\ket{1} $, using the \textit{Dirac} notation for the base vectors $\ket{0} = \left(1,0\right)^\intercal$ and $\ket{1}= \left(0,1\right)^\intercal$ of each individual qubit \cite{Nielsen2010}.
\smallskip

The physical state~$y$ is represented variably by the control vector $\upb^l \in \mathbb{R}^{c+1},  \upb^l: = \big(\lambda_0^l, \lamb_c^l\big)^\intercal$ ($c \in \mathbb{N}$) and realized by the time-dependent ansatz 
\begin{equation}
    y(x,t^l, \upb^l) = \lambda_{0}^{l} \, u(x, {\lamb_c^l}) \, , 
    \label{eq:Generic_Ansatz}
\end{equation}
which represents the interface for the exchange of classical/quantum mechanical information.
Due to the probabilistic nature of quantum states, one needs to enforce the \textit{partition of unity} constraint in the $L^2$ inner product of the ansatz function~$u$~\cite{Troutman1996,Werner2005}, viz. $\big(u^*, u\big)_{L^2} = 1$. Here, the asterisk represents the complex conjugate of the quantum states, and the factor~$\lambda_0^l$ is introduced to account for general solutions not only norm-preserving ones. This enables a discrete space representation of $u\big(x,\lamb_c^l\big)$, viz. vector~$\ub$, by the amplitudes $u_k$ of the quantum state~$\ket{u}$ in the quantum register as $\ub = \sum\limits_{k} u_k \eb_{k-1} = \ket{u}$ (known as \textit{amplitude encoding}). Mind that, contrary to real-valued physical states~$y$, quantum amplitudes~$u_k$ include phase information and thus employ a complex notation, and that the quantum register is in the following restricted to the interior points, viz.
\begin{equation}
     \ub = \Bigg(u_0, \Big( \, \underbrace{\sum\limits_{k=1}^{N_\text{p}} u_k \eb_{k-1}}_\text{register}\, \Big)^\intercal, u_{N_\text{p}+1} \Bigg)^\intercal \, . 
     \label{eq:discretization_register}
\end{equation}

\subsection{Hybrid Classical-Quantum Approach} 
\label{sec:hybrid}
For each time step~$l$, the \textsc{VQA} approach minimizes the discretized cost function~\eqref{eq:Generic_Opt_Approx}, by adjusting the control parameters $\upb^l: = \big(\lambda_0^l, \lamb_c^l\big)^\intercal$ that shape $\lambda_0^l\,\ub^l$. This quantum state~$\ket{u^l}$ results from the parameterized quantum gate $U\big(\lamb_c^l\big)$ applied to $\ket{0}$.

The optimization is executed on a hybrid classical-\textsc{QC} hardware, as indicated in \Fig{fig:optimizer}. On the right-hand side of \Fig{fig:optimizer}, quantum registers are used to evaluate the cost function ${J}$ on the \textsc{QC}. At the same time, the classical computer updates the control $\upb_i$, within an optimization procedure depicted on the left-hand side of \Fig{fig:optimizer}. 
\begin{figure}[htbp]
    \centering
    % Define block styles
\tikzexternaldisable
\tikzstyle{decision} = [diamond, draw, text width=5em, text badly centered, node distance=3cm, inner sep=0pt]
\tikzstyle{decisionklein} = [diamond, draw, text width=3.5em, text badly centered, node distance=3cm, inner sep=0pt]
\tikzstyle{decisiongross} = [diamond, draw, text width=5em, text badly centered, node distance=3cm, inner sep=0pt]
\tikzstyle{block} = [rectangle, draw, text width=10em, text centered,  minimum height=4em]
\tikzstyle{blockklein} = [rectangle, draw, text width=6em, text centered,  minimum height=4em]
\tikzstyle{blockgross} = [rectangle, draw, text width=8em, text centered,  minimum height=4em]
\tikzstyle{blockmitte} = [rectangle, draw, text width=9em, text centered,  minimum height=4em]
\tikzstyle{blockmitteclean} = [rectangle, text width=8em, text centered,  minimum height=4em]

\tikzstyle{quantum} = [rectangle, draw, text width=18em, text centered,  minimum height=2em, fill=gray]
\tikzstyle{classical} = [rectangle, draw, text width=8em, text centered,  minimum height=2em]
\tikzstyle{exit} = [rectangle, draw, text width=5em, text centered,  minimum height=4em]
\tikzstyle{blockk} = [rectangle, draw=blue, text width=7em, text centered, rounded corners, minimum height=4em] 
\tikzstyle{cloud} = [draw, ellipse, node distance=4cm,minimum height=2em]
\tikzstyle{line} = [draw]
\tikzstyle{arrow} = [draw, -latex] 
\tikzstyle{stopblock} = [rectangle, minimum width=1cm, text width=3.5cm, minimum height=1cm,text centered]
\tikzstyle{startblock} = [rectangle, minimum width=1cm, text width=3.5cm, minimum height=1cm,text centered]
  
\begin{tikzpicture}[node distance = 1.5cm, scale=1]
%% Outer nodes
    \node [blockmitte] (opt) {$\min\limits_{\upb^{l}}J^{l}\left(\upb^{l}\right)$};
    \node[startblock, left of= opt,xshift=-2cm](start){ \textbf{start}\;$\bullet$};
    \node[stopblock, right of=opt,xshift=+1.5cm](stop){$\bullet$\; \textbf{stop}};

%%Shape nodes
    \node [blockklein, below of= opt,xshift=-3cm,yshift=-0.4cm, node distance = 3.2cm](co1){Evaluate the Objective $J(\lambda_0, \lamb_c)$ \eqref{eq:Generic_Opt_Approx}}; 
    \node [decisionklein, below of=co1,yshift=+1cm] (co2){conv.?};
    \node[blockgross, below of= co2, yshift=-0.5cm] (co3){Update\\
    $\upb_{i+1} = \upb_{i} + \Delta\upb$};

%% QC Nodes
    \node [quantum, right of=co1, xshift=5.6cm,yshift=-1.7cm](qc1){Bilinear:\\
    \begin{quantikz}[column sep=0.2cm,row sep=0.2cm]
    \lstick{$\ket{0}$}& \gate{H} &  &\phase{^\text{CP}}\wire[d][1]{q}& \gate{H} & & \meter{\to B} \\ 
    \lstick{$\ket{0}^{\otimes n}$}& \gate[style={fill=darkgray}]{U(\lamb_{c_i})} & \wire[r][1]["\text{IP}"{above,pos=0.1}]{q}& \gate[style={fill=lightgray}]{\rotatebox{90}{\text{QNPU}}} &\wireoverride{n}\wire[l][1]["\text{OP}"{above,pos=0.6}]{q} &  & \\
    \end{quantikz}\\[4mm]
    Linear:\\
     \begin{quantikz}[column sep=0.2cm,row sep=0.2cm]
    \lstick{$\ket{0}$}& \gate{H} &\phase{^\text{CP}}\wire[d][1]{q}&  &\phase{^\text{CP}}\wire[d][1]{q}& \gate{H} & & \meter{\to F} \\ 
    \lstick{$\ket{0}^{\otimes n}$}& &\gate[style={fill=darkgray}]{U(\lamb_{c_i})} & \wire[r][1]["\text{IP}"{above,pos=0.1}]{q}& \gate[style={fill=lightgray}]{\rotatebox{90}{\text{QNPU}}} &\wireoverride{n}\wire[l][1]["\text{OP}"{above,pos=0.6}]{q}  &  & \\
    \end{quantikz}
     };

%% Boxes
\node[draw,inner xsep=4mm, inner ysep=2.5mm, dashed,overlay, label=below : Optimization ,fit=(co1) (co2) (co3) (qc1) ] (Optimization_Box) {};
\node [classical, below of=co2, xshift=0cm, yshift=-3.5cm ](CDevice){\textbf{Classical Device}};
\node [quantum, right of= CDevice, xshift=5.6cm](QDevice){\textbf{Quantum Device}};

%% PATHs
\path [arrow] ([xshift=-1.3cm]start.east) -- (opt.west);
\path [arrow] (opt.east) -- ([xshift=+1.3cm]stop.west);
\path [arrow] ([xshift=-0.1cm]opt.south) |- node [anchor=east] {} ([xshift=+1.3cm, yshift=-1cm]start.south) -| (co1.north);
\path [arrow] ([yshift=0.2cm]co1.east) -- node [anchor=south] {$\lamb_{{c}_i}$} ([yshift=1.9cm]qc1.west);
\path [arrow] ([yshift=1.5cm]qc1.west) --node [anchor=north] {$J_i$} ([yshift=-0.2cm]co1.east);
\path[arrow] (co1.south) -- (co2.north);
\path[arrow] (co2.east) -| node [anchor=north] {yes} ([xshift=+0.1cm]opt.south);
\path[arrow] (co2.south) --node [anchor=east] {no}  (co3.north);
\path[arrow] (co3.west) -| ++(-0.55,0) -- ++(0,2) -- ++(0,0.55)
|- node[xshift=+0.1cm, yshift=-2.0cm, text width=2.5cm,anchor=west] {$i+1$}(co1.west);
\end{tikzpicture}
\tikzexternalenable
    \caption{Sketch of the hybrid classical-\textsc{QC} optimization in the \textsc{VQA}. The bilinear and linear circuit on the right-hand side show, the Hadamard gate~$\pmb{H}$, the ansatz gate $\pmb{U\big(\lamb_c\big)}$, the \textsc{QNPU} module, and the measurement gauge on the ancilla. The abbreviation \text{CP} marks a control port while \text{IP} \& \text{OP} denote input/output ports, respectively.}
    \label{fig:optimizer}
\end{figure}
As outlined in the introduction to this section, the readout process is reduced to a single qubit evaluation to mitigate the measurement problem on \textsc{QC}s \cite{Endo2020,Guseynov2023}. This scalarization of the \textsc{PDE} problem is achieved through a Hadamard test-based \textsc{VQA} that delivers only scalar quantities of interest, such as the cost function derived from the weighted residual formulation of the \textsc{PDE} in \Eq{eq:WResidualEq}, instead of complete physical fields. Therefore, both quantum circuits representing the bilinear and linear integrals of \Eq{eq:Generic_Opt_Approx}, on the right-hand side of \Fig{fig:optimizer}, are based on the Hadamard test \cite{Nielsen2010} with an upper measurement qubit (ancilla) and the lower working qubits, all initialized to the base state~$\ket{0}$. The only difference between the two circuits of \Fig{fig:optimizer} is the ancilla-controlled ansatz gate~$U(\lamb_c^l)$ (bottom circuit), which corresponds to the overlap operation (i.e., a scalar product) required in the linear term~$F$.

For the sake of completeness, the employed parameterization~$U(\lamb_c^l)$ of the quantum state $\ket{u^l} = U(\lamb_c^l) \ket{0}$ should also be mentioned here. We apply a real-valued ansatz for real-valued solutions suggested by \citeauthor{BravoPrieto2023}~\cite{BravoPrieto2023}, which is also used in other studies \cite{Sato2021,Leong2022} related to our research. It consists of parameterized $R_\text{Y}(\lambda_j)$ gates and increases the expressiveness by numerous controlled \textsc{Z}-gates. The chosen ansatz is depicted in Fig.~\ref{fig:ansatz} and is characterized by linear scaling in the number of trainable parameters to reduce the related complexity. The fixed structure, shown in Fig.~\ref{fig:ansatz}, is based on a hardware-efficient bricklayer topology \cite{Nakaji2021} and supports arbitrary ansatz types with variable depth $d$ to increase the expressiveness of the solution space, cf.~\cite{Over2024}. The complexity of the ansatz is addressed in Sec.~\ref{sec:qcfd:gatecomplexity}. However, a detailed analysis of the used ansatz is not intended here.
\begin{figure}[htbp]
    \centering
    \include{Ansatz}
    \caption{Employed brick-layer ansatz as proposed in Ref.~\cite{BravoPrieto2023}. Each $\pmb{R_\text{Y}=exp{(-\lambda_
    j\, i\, \textsc{Y}/2)}}$ gate is parameterized by a unique control parameter $\pmb{\lambda_j}$ of the control vector $\pmb{\lamb_c}$ and $\pmb{i}$ is the imaginary unit applied to the \textsc{Y}-gate.}
    \label{fig:ansatz}
\end{figure}

\subsection{Building Blocks for Physical Modeling with \textsc{QNPU}s} \label{sec:qnpus}
The \textsc{QNPU} block in \Fig{fig:optimizer} indicates the part where the computational or physical modeling is implemented. To this end, each summand in~$J$, given in \Eq{eq:WResidualEq}, is translated into quantum gates, resulting in an individual \textsc{QNPU}. Similarly, boundary conditions also contribute to~$J$ through their particular \textsc{QNPU}s, cf. \citeauthor{Over2024}~\cite{Over2024} for details. 

Prior to discussing the modeling of the individual spatial discretization schemes with quantum gates in the Sec.~\ref{sec:QCFD}, we explain the building blocks required for this task. The description opens with the external contributions of a potential field $p(x_k,t^l)$, e.g., used to represent inhomogeneous material properties, followed by permutation operators, so-called adder circuits, to compose FD expressions, and finishes with the overlap operation for source-term contributions. All the quantum circuits are exemplified for $n=4$ qubits, and their general patterns are presented. The circuits are labeled with an Input Port (\textsc{IP}) entered by the state~$\ket{u}$, a Control Port (\textsc{CP}) applied to the ancilla (top wire) depicted in the circuits of Fig.~\ref{fig:optimizer}, and an Output Port (\textsc{OP}) recovering a mixed state. The latter is indicated by~$(\ast)$ as the output state can not be written as a product state. To motivate the three building blocks below, we refer to the example of a simple first-order \textsc{FD} approximation for the one-dimensional expression $y_k \, a_k \, {\partial y_k}/{\partial x}$ in a location $x_k$ by $y_k a_k (y_{k+1} - y_{k})/\Delta x$ which involves products between unknowns at the same ($k,k$) and at different locations ($k$, $k+1$). 
\smallskip

\subsubsection{Potential Term} \label{sec:x1}
The term $p(x_k, t^l)$ models contributions of potentials in the general form $\sum\limits_{k=1}^{N_\text{p}} \left[ \lambda_0^2 u_k^*p_k u_k \right]^l $. 
With reference to \Eq{eq:WResidualEq} this obviously involves parts of the bilinear term~$B(y^l,y^l)$ evaluated for the current time step, i.e., $\int y^l \left( a_5 + \frac{a_1}{\Delta t^2} + \frac{a_2}{\Delta t}\right)y^l \,dx$. Depending on the employed spatial discretization with an \textsc{FD} scheme, the potential term also includes a contribution from the diffusive term $\int y^l\frac{\partial}{\partial x} \big(a_3 \frac{\partial y^l}{\partial x}\big)\,dx$. Moreover, the nonlinear convective term $\int y^{l} a_4 \big(\frac{\partial y^{l-1}}{\partial x}\big)\,dx$ also contributes to $p$ for an explicit treatment of the flux coefficient $a_4=y^{l-1}$ as it is also indicated in \Fig{fig:qnpu:potential_potential_adder}. 

The corresponding \textsc{QNPU} circuit is assembled by $n$~Toffoli gates and $n$~additional auxiliary qubits because the product $u_k^* p_k u_k$ is non-unitary \cite{Sarma2023}, see~\Fig{fig:qnpu:potential}. Note that the potential function~$p$ must also be represented as a quantum state. To this end, state tomography can be applied for training the ansatz $P(\bar{\lamb}_c)$ parameterizing $\ket{p}=P(\bar{\lamb}_c)\ket{0}$. Alternatively, one can employ a direct representation with a series of quantum gates, using tensor quantum programming~\cite{Ran2020}. The present study utilizes a hardware efficient implementation using the ansatz gate $U(\bar{\lamb}_c)$ for encoding~$p$. The latter is subject to an additional optimization problem to determine the trained control vector $\bar{\lamb}_c$ from 
\begin{equation}
     \bar{\lamb}_c =  \min\limits_{\lamb_c} \big(1 -\lambda_0\braket{u(\lamb_c)|p}\big) \quad  \text{ with } \frac{1}{\lambda_0^2}=\sum \limits_1^{N_\text{p}} p_k^* p_k \, . 
    \label{eq:OptAnsatz}
\end{equation}
To differentiate the ansatz of a potential~$p$, $P(\bar{\lamb}_c)$ indicates the application of the trained parameters~$\bar{\lamb}_c$ to $U(\bar{\lamb}_c)$. Multiple potentials can be staggered to create higher-order pointwise products, as required for nonlinearities~\cite{Lubasch2020}.
\begin{figure}[htbp]
    \centering
    \input{F_Potential}
    \caption{\textsc{QNPU} schematic for the potential contribution $\pmb{P}$, e.g., for $\pmb{n=4}$ qubits. The ansatz gate for encoding $\pmb{p}$ is marked with $\pmb{P(\bar{\lamb}_c)}$, where  $\pmb{\bar{\lamb}_c}$ are the trained parameters. The control ports are labeled \text{CP}, the input ports are labeled \text{IP}, and the output ports are labeled \text{OP}.}
    \label{fig:qnpu:potential}
\end{figure}
\smallskip

\subsubsection{Adder Circuits} \label{sec:x2}
Adders are used to perform classical bit shift operations \cite{Nielsen2010,Vedral1996,Camps2024} required for products of shifted locations, e.g., $u_k^{*l} u_{k+1}^l$ or $u_k^{*l} u_{k-1}^l$. The adder circuit is the building block to include any \textsc{FD} stencil in the weighted residual approach forming the cost function~$J$ and they are key circuits for the reduction of the gate complexity, see Sec.~\ref{sec:qcfd:gatecomplexity}. For the example of a weighted integration of a forward differentiation scheme $y \, \partial y/\partial x \sim y_k (y_{k+1}-y_{k})$, the adder circuit performs (shift-down) \cite{Lubasch2020}, 
\begin{equation}
    \label{eq:Laplace}
    \sum\limits_{k=1}^{N_\text{p}}u_k \ket{\text{bin}\left(k-1\right)} \underset{k\neq N_\text{p}}{\to}  \sum \limits_{k=1}^{N_\text{p}} u_{k+1}  \ket{\text{bin}\left(k-1\right)}\,. 
\end{equation}
Mind that for $k = N_\text{p}$ the periodicity of the quantum register needs to be respected, following the modulo operation ${(k \mod N_\text{p})+1}$.

A half-adder circuit with one \textsc{CNOT} and one Toffoli gate is applicable for $n=2$ qubits \cite{Nielsen2010}. In the case of $n > 2$, the additional ($n-2$) carry (auxiliary) qubits lead to a full-adder circuit with $n-2$~\textsc{CNOT}s and $2n-2$~Toffoli gates. Figure~\ref{fig:qnpu:adder} depicts the corresponding \textsc{QNPU} circuits for $n=4$~qubits. 
\begin{figure}[htbp]
    \centering
    \begin{subfigure}[t]{0.48\textwidth}
          \input{F_Adder}
          \caption{Shift-down adder $\pmb{A}$, cf. \cite{Lubasch2020}.}
          \label{fig:qnpu:adder:down}
    \end{subfigure}
    \begin{subfigure}[t]{0.48\textwidth}
          \input{F_Adder_up}
          \caption{Shift-up (adjoint) adder $\pmb{A^\dagger}$.}
          \label{fig:qnpu:adder:up}
    \end{subfigure}   
    \caption{\textsc{QNPU} schematic for the adder (shift) circuits, e.g., for $\pmb{n=4}$ qubits. The control ports are labeled \text{CP}, the input ports are labeled \text{IP}, and the output ports are labeled \text{OP}.}
    \label{fig:qnpu:adder}
\end{figure}

Two adder circuits are required, one for assessing the upper diagonals (e.g., a forward differencing scheme) and one for assessing the lower diagonals (e.g., a backward differencing scheme). The adder circuit~$\Ab$ displayed in \Fig{fig:qnpu:adder:down} performs a shift-down operation (forward differencing) and its conjugate-transpose~$\Ab^{\dagger}$ in \Fig{fig:qnpu:adder:up} performs a shift-up (backward differencing) instead. Furthermore, multiple shift-up or shift-down operations can be stacked sequentially by applying the circuit $q$~times. This provides the possibility of representing arbitrary \textsc{FD} stencils, as detailed in Sec.~\ref{sec:QCFD}. 
Mind that depending on the capabilities of the \textsc{QC} hardware, the number of involved gates can be reduced by using multi-controlled gates when these are native on the particular hardware~\cite{McDonnell2022}. For further discussion, the reader is referred to the section on complexity in Ref.~\cite{Over2024}. 
\smallskip

\subsubsection{Source Term} \label{sec:x3}
The contribution $\int y f\, dx$ to $J$ in \Eq{eq:WResidualEq} accounts for the source term and is represented as an overlap operation (scalar product) on a \textsc{QC}. Therefore, the linear term in \Fig{fig:optimizer} has a ``controlled'' $U(\lamb_{c_i})$ gate in contrast to the bilinear term. This circuit is used to realize products such as $u_k^{*l}u_k^{l-1}$, where the \textsc{QNPU} is realized with the trained approach $U^\dagger(\lamb^{l-1})$.

\medskip 
All three fundamental building blocks described above, i.e., potential, adder, and source \textsc{QNPU}s, can be combined to model discretized transport operators. For example, the combination of potential and shift \textsc{QNPU}s is used to implement an advective contribution to the objective functional, i.e., $\int y\, a_4 \big(\frac{\partial y}{\partial x}\big)\,dx$. This combination is illustrated in \Fig{fig:qnpu:potential_adder}. Herein, the potential $P(\bar\lamb_{c})$ is trained to represent the coefficient distribution $a_4$.
\begin{figure}[htbp]
    \centering
    \input{F_Potential_Adder}
    \caption{\textsc{QNPU} schematic for the adder-potential $\pmb{{A}_p}$ for the example of $\pmb{n=4}$ qubits. The ansatz gate for the potential~$\pmb{p}$ is indicated by $\pmb{P(\bar{\lamb}_c)}$. The abbreviations \textsc{CP}, \textsc{IP}, \textsc{OP} mark the control, input, and output ports, respectively.}
    \label{fig:qnpu:potential_adder}
\end{figure}

Figure~\ref{fig:qnpu:potential_potential_adder} depicts a \textsc{QNPU} implementation of a nonlinear convective contribution to the objective functional, i.e., $\int y^l a_4^{l-1} \big(\frac{\partial y^{l-1}}{\partial x}\big)\,dx$, by the combination of two potentials (\Fig{fig:qnpu:potential}), one adder (\Fig{fig:qnpu:adder}), and one source term. To build the convective term following an explicit treatment, the coefficient is $a_4^{l-1}=p\;y^{l-1}$ and thus the old control variable~$\lamb_c^{l-1}$ is used for $U^\dagger(\lamb_c^{l-1})$ in \Fig{fig:qnpu:potential_potential_adder}. The trained ansatz $P(\bar\lamb_{c})$ represents a potential~$p$, which can be used to imply a pointwise modulation of the coefficient, e.g., being zero or one. The adder as depicted here implements a forward \textsc{FD} approximation and can be modified easily to implement other (higher-order) \textsc{FD} approximations. The \textsc{QNPU} is subsequently realized in combination with another $U^\dagger(\lamb^{l-1}_c)$, substituted in the source term, cf. Sec.~\ref{sec:x3}, to prepare the state $a_4^{l-1} \big(\frac{\partial y^{l-1}}{\partial x}\big)$. The assembled \textsc{QNPU} in \Fig{fig:qnpu:potential_potential_adder} is used in the bottom circuit on the right-hand side of \Fig{fig:optimizer}, which in turn is multiplied with the current state to get the final cost function contribution, i.e., $\int y^{l} \left[ ... \right]\,dx$. 
\begin{figure}[htbp]
    \centering
    \input{F_Potential_Potential_Adder}
    \caption{\textsc{QNPU} schematic for the adder-potential-potential $\pmb{{A^\dagger}_{p^\dagger p^\dagger}}$ with the ansatz $\pmb{U^\dagger(\lamb^{l-1}_c)}$ for the example of $\pmb{n=4}$ qubits. The ansatz gates to model the convective coefficient $\pmb{a_4}$ are the potential $\pmb{p}$, indicated with $\pmb{P^\dagger(\bar{\lamb}_c)}$, and the ansatz gate $\pmb{U^\dagger({\lamb^{l-1}_c})}$ (nonlinearity). The subsequent gate $\pmb{U^\dagger(\lamb^{l-1}_c)}$ represents the state at the previous $\pmb{l-1}$ (explicit scheme) time step. The abbreviations \textsc{CP}, \textsc{IP}, \textsc{OP} mark the control, input, and output ports, respectively.}
    \label{fig:qnpu:potential_potential_adder}
\end{figure}

%###########################################
%       QCFD
%###########################################
\section{Quantum Computational Fluid Dynamics}
\label{sec:QCFD}
Analogous to classical \textsc{CFD}, the presented \textsc{QCFD} module is divided into pre-processing, computation (\textsc{VQA}), and post-processing steps~\cite{Ferziger2020}. During pre-processing, initial conditions, mask functions, and source terms are encoded in the quantum register. This step can be implemented variationally on quantum-classical hardware and requires additional optimization problems to be solved. For the execution of the introduced \textsc{VQA}, the current manuscript employs a library of standard \textsc{CFD} discretization schemes and boundary conditions~\cite{Over2024}. The corresponding quantum circuits presented in Sec.~\ref{sec:Quantum} are applied to the discretized problem following the concepts presented in this section. The final post-processing step can also be integrated into the proposed \textsc{VQA}~\cite{Lubasch2020}, enabling efficient on-the-fly evaluation of quantities of interest such as lift and drag coefficients. Bear in mind that the ansatz can also be post-evaluated by saving the final ansatz parameters after each optimization step. This supports recovering the field solutions for a detailed analysis on either classical or hybrid hardware.

The current section describes the translation of \textsc{FD} stencils into the quantum framework. Due to the amplitude encoding strategy, i.e.,~\Eq{eq:discretization_register}, its application is limited to global ansatz functions. The technique is first demonstrated for elementary \textsc{FD} stencils (Sec.~\ref{sec:qcfd:diffschemes}) and subsequently generalized in Sec.~\ref{sec:qcfd:general}. Thereafter, we employ the general strategy for modeling an inhomogeneous thermal diffusivity (Sec.~\ref{sec:qcfd:material}) as well as for higher-order convection schemes (Sec.~\ref{sec:qcfd:convection}).

\subsection{Elementary FD Schemes}
\label{sec:qcfd:diffschemes}
Using a midpoint integration rule, the discrete cost function contributions for approximating first derivatives by elementary \textsc{FD} formulae on an equidistant mesh follow from 
\begin{align}
    &\text{1st-order forward:} \hphantom{----------} \int_\Omega y\frac{\partial y}{ \partial x} dx \approx   \sum_k y_k \left(y_{k+1}-y_{k}\right)=J_\text{forw-1} \,, \label{eq:forward}\\
    &\text{1st-order backward:} \hphantom{-----} \qquad \quad \; \int_\Omega y \frac{\partial y}{ \partial x}  dx \approx \sum_k   y_k \left(y_k-y_{k-1}\right)= J_\text{back-1} \,, \label{eq:backward} \\
    &\text{2nd-order central:} 
   \hphantom{--------} 
    \int_\Omega y\frac{\partial y}{ \partial x} dx \approx  \frac{1}{2} \sum_k y_k \left(y_{k+1}-y_{k-1}\right)=J_\text{cent-1}\,. \label{eq:central} 
    \end{align}
Similarly, the approximate \textsc{FD} contribution that involves a second derivative is of 
    \begin{equation}
        \text{2nd-order central:} \hphantom{FD}  \quad \; \int_\Omega y\frac{\partial^2 y}{ \partial x^2} dx \approx \frac{1}{\Delta x} \sum_k y_k \left(y_{k+1}-2y_{k}+y_{k-1}\right)=J_\text{cent-2}\, .
        \label{eq:central2}
    \end{equation}

Substituting the quantum ansatz of Eqns.~(\ref{eq:Generic_Ansatz},~\ref{eq:discretization_register}) into the Eqns.~(\ref{eq:forward},~\ref{eq:backward}), the cost function contributions read
\begin{align}
     J_\text{forw-1} &= 
     \lambda_0^2 \Big(\sum_k   u_k^* (\lamb_c)u_{k+1}(\lamb_c)- 1 \Big) \,,\label{eq:forward_qc}\\
       J_\text{back-1} &= \; 
    \lambda_0^2 \Big(1 - \sum_k   u_k^* (\lamb_c)u_{k-1}(\lamb_c)\Big) \,. \label{eq:backward_qc}
\end{align}

The \textsc{QNPU} of \Fig{fig:qnpu:adder} applies the adder~$\Ab$ to implement $ \sum_k  u_k^* (\lamb_c)u_{k+1}(\lamb_c)$ for the forward differencing, while the backward difference term, $\sum_k u_k^* (\lamb_c)u_{k-1}(\lamb_c)$, is modeled by~$\Ab^\dagger$ instead. 

Analogous to classical \textsc{FD}, the quantum implementation of the central difference in \Eq{eq:central} can be expressed by the averaging of forward~\eqref{eq:forward} and backward~\eqref{eq:backward} stencils. However, when introduced to the weighted residual method, the result vanishes for a periodic case, since the sums of the forward~\eqref{eq:forward_qc} and backward~\eqref{eq:backward_qc} differences neutralize, i.e.,  $J_\text{cent-1}$ does not contribute to~$J$. Similarly, the \textsc{CDS} approximation of the second derivative $J_\text{cent-2}$, could be expressed as $J_\text{cent-2}=(J_\text{for-1}-J_\text{back-1})/\Delta x$. The approximation results in a symmetric tridiagonal (Toeplitz) matrix, which is represented by the adder circuit of \Fig{fig:qnpu:adder} and can be efficiently executed on a \textsc{QC}~\cite{Over2024,Lubasch2020}. In a periodic case, the difference of the two sums agrees with twice its components and the cost function contribution simplifies to 
\begin{equation}
    \label{eq:seconddiff_quantum}
    J_\text{cent-2} = +2/(\Delta x) J_\text{for-1}=-2/(\Delta x) J_\text{back-1} \, .
\end{equation}

\subsection{Band Matrix Operations with \textsc{QNPU}s}
\label{sec:qcfd:general}
We now generalize the formalism of Sec.~\ref{sec:qcfd:diffschemes} to the diagonals of band matrices~\cite{Camps2024}, which typically occur for \textsc{FD} approximations of transport equations on Cartesian grids. Since the implementation of non-periodic boundary conditions is already covered in Ref.~\cite{Over2024} and will not be repeated here, we again restrict the derivation to periodic boundaries.

The notation of a band matrix~$\Cb$, with bandwidth~$b_\text{w}$, obtained from a \textsc{FD} discretization reads $\Cb=c_{kj} \, \eb_k \otimes \eb_j\,.$ A~vector representation of the band matrix's $q$-th non-zero diagonal, where $q \in \left[-\lfloor(b_\text{w}-1)/2\rfloor,+\lfloor(b_\text{w}-1)/2\rfloor\right]$ ($\lfloor \dots \rfloor$ -- integer operation), is written as
$\pb_q = p_{q_k} \eb_k = c_{ \underline{k}(j+q) }  \,  \delta_{ \underline{k}j } \,  \eb_k$. Here, $\delta_{\underline{k}j}$ denotes the unit tensor where the underline marks a non-contractable index. The operation selects the respective $q$-th diagonal through the index $(j+q)$ when contracted by the dot product. Thus, $q=0$~points to the main diagonal, and $q>0$ or $q<0$ denotes the $q$-th super- or subdiagonals, respectively. Note that all diagonals are of the same length in a periodic setting.
The action of the matrix~$\Cb$ on the cost function~$J$ is separated into (non-zero) diagonal contributions as 
\begin{equation}
    \sum_o^{\min{(q)},\max{(q)}} {J}_o = \sum_{k} y_k p_{o_k} y_{(k+|o|)} \Delta x \,. 
    \label{eq:qc:J_diag_contributions}
\end{equation}
In an analogy to the variable potential term~$p$, cf. Sec.~\ref{sec:x1}, the ansatz gate $U(\lamb_c)$ is trained to reproduce the non-constant $q$-th diagonal of $\Cb$ by solving an overlap problem (cf. \Eq{eq:OptAnsatz}) before its use in the corresponding \textsc{QNPU}. The normalization conditions for the description of a vector on a \textsc{QC} need to apply for each non-zero diagonal of the band matrix $\Cb$. For this purpose, the scaling factor ${\tilde \lambda_q}$, which corresponds to the factor ${\lambda_0}^l$ in \Eq{eq:Generic_Ansatz}, is introduced. As a result, the vector of the $q$-th diagonal $\pb_q$ is encoded by $\tilde\lambda_q \tilde \pb_q$. In the case of Toeplitz matrices, i.e., constant diagonal entries, the representation of the diagonal by the ansatz $U(\lamb_c)$ can be expressed, up to normalization, by Hadamard gates, avoiding the solution of the optimization problem of \Eq{eq:OptAnsatz}. 

The contribution of the main diagonal ($q=0$) to the cost function of \Eq{eq:qc:J_diag_contributions} is formed by the normalized diagonal entries, with the state $\tilde\pb_0$. The evaluation for this diagonal is conducted by the circuit of \Fig{fig:qnpu:potential} and reads 
\begin{equation}
         J_0 = \lambda_0^2  \sum_k u_k^* \underbrace{{\tilde \lambda_0} \tilde{p}_{0_k}}_{p_{0_k} = {c}_{kk}} u_k \Delta x\,. 
        \label{eq:qc:main_diagonal_contribution}
\end{equation}
In case $c_{kk}\equiv \text{const.}$, the circuit evaluation is unnecessary, and the cost function contribution results in $J_0/\Delta x=\lambda_0^2 {\tilde\lambda_0} \text{, e.g., } \lambda_0^2 c_{00} $. For the off-diagonals $|q|\geq 1$, the same formalism is implemented, with the diagonals represented by a quantum state. Here, the \textsc{QNPU} in \Fig{fig:qnpu:potential_adder} is applied, with the adder operator appearing $q$~times. As an example, the cost function contributions for 
$q=\pm1$ are
\begin{equation}
    {J}_{1}= \lambda_0^2  {\tilde\lambda_1}\sum_k u_k^*  \tilde{p}_{1_k} ({\Ab} \ub)_k \Delta x\,,
\quad \text{and} \quad  
    {J}_{-1}= \lambda_0^2  {\tilde\lambda_{-1}}\sum_k u_k^*  \tilde{p}_{{-1}_k} (\Ab^{{\dagger}} \ub)_k \Delta x\,,
    \label{eq:qc:diagonals_contribution}
\end{equation}
where the conjugate-transpose~${\Ab}^{\dagger}$ performs a shift-up instead of a shift-down, cf.~Sec.~\ref{sec:qnpus}. 
Although the presented approach is applicable to every matrix, it is most suitable for symmetric matrices because the off-diagonal entries are identical for the same $|q|\geq 1$, and the cost function contributions for $q<0$ can be dropped by doubling the contributions of $q>0$. Applying this to the cost function contributions for $q=\pm 1$ yields
\begin{equation}
    {J}_1 + {J}_{-1}= 2 \lambda_0^2  {\tilde\lambda_1} \sum_k u_k^* {p}_{1_k} ({\Ab}\ub)_k \Delta x \,.
    \label{eq:generalCDS2} 
\end{equation}
Note that \Eq{eq:generalCDS2} is a generalization of the discussion in Sec.~\ref{sec:qcfd:diffschemes}. As a last remark, it is important to highlight that training a quantum state allows boundary conditions to be introduced by manipulating the corresponding entries in~$\pb_q$. 

\subsection{Non-Constant Diffusivity}
\label{sec:qcfd:material}
Inhomogeneous material properties, such as a spatially variable thermal diffusivity~$\alpha(x)$, frequently occur in engineering applications. To illustrate the related quantum model, a periodic diffusion operator in a weighted residual formulation is considered, cf.~\Eq{eq:WResidualEq},   
\begin{equation}
\int_\Omega y(x) \frac{\partial}{\partial x} \left(\alpha(x) \frac{\partial }{\partial x}y(x,t) \right)  dx\, .  
\label{eq:material}
\end{equation}
The derivatives in \Eq{eq:material} are approximated by a consecutive application of backward (outer) and forward (inner) \textsc{FD}s, resulting in the following \textsc{FD} matrix for the interior points
\begin{equation}
\begin{split}
        & \frac{1}{{\Delta x}^2}
    \underbrace{\begin{bmatrix}
          1 &  0     &  \dots  &-1 \\
         -1 &  1     &  \dots  & 0 \\
            & \ddots &  \ddots &   \\
          0 & \dots  & -1      & 1 \\
     \end{bmatrix}}_{\text{backward (outer) \textsc{FD}}}
     \begin{bmatrix}
         \alpha_1 & 0        & \dots  & 0 \\
          0       & \alpha_2 & \dots  & 0 \\
                  &          & \ddots &   \\
          0       & \dots    & 0      & \alpha_{N_{\text{p}}}
     \end{bmatrix}
     \underbrace{\begin{bmatrix}
         -1 &      1 &  \dots &  0 \\
            &     -1 &      1 &    \\
            &        & \ddots &  \ddots \\
         1  &  \dots &  0     & -1 
     \end{bmatrix}}_{\text{forward (inner) \textsc{FD}}}
     \begin{bmatrix}
         y_1\\
         y_2\\
         \vdots\\
         y_{N_\text{p}}
     \end{bmatrix}\, \\
     &= \frac{1}{{\Delta x}^2}
     \begin{bmatrix}
         -(\alpha_{N_\text{p}} + \alpha_1) & \alpha_1           & \dots        & \alpha_{N_\text{p}} \\
         \alpha_1             & -(\alpha_1+\alpha_2) &\alpha_2      & 0 \\
                              & \ddots             & \ddots       & \ddots \\
         \alpha_{{N_\text{p}}}             & \dots              & \alpha_{N_\text{p}-1} & -(\alpha_{N_\text{p}-1} +\alpha_{N_\text{p}})
    \end{bmatrix}
     \begin{bmatrix}
         y_1\\
         y_2\\
         \vdots\\
         y_{N_\text{p}}
    \end{bmatrix} \approx \frac{\partial}{\partial x} \left(\alpha(x) \frac{\partial }{\partial x} y(x,t) \right) \, .
    \label{eq:diff_noconst_matrix}
    \end{split}
\end{equation}

From \Eq{eq:diff_noconst_matrix}, it is evident that a symmetric band matrix is obtained. As elaborated in Sec.~\ref{sec:QCFD}, this allows modeling the dynamics using only $\lfloor b_\text{w}/2 \rfloor+1 = 2$ quantum circuits. This observation is confirmed by forming the discretized expression of \Eq{eq:material}, which reads 
\begin{equation}
  J_\alpha = \frac{1}{\Delta x^2} \Big(\sum_k y_k \underbrace{\left(-\alpha_{k}-\alpha_{k-1}\right)}_\text{diagonal}y_k + 2\sum_k y_k \underbrace{\alpha_{k}}_{\substack{\text{super-/} \\ \text{sub-} \\ \text{diagonal}}} y_{k+1}\Big)\Delta x \, .  
\end{equation}
To show the similarity with the general framework presented in Eqns.~(\ref{eq:qc:main_diagonal_contribution}-\ref{eq:qc:diagonals_contribution}), we define ${c}_{kk}=\left(\alpha_k+\alpha_{k-1}\right)=\tilde\lambda_0 \tilde p_{0_k}$ and ${c}_{k(j+1)}={c}_{k(j-1)}=\alpha_k=\tilde\lambda_1\tilde p_{1_k}$. The resulting cost function contribution for the quantum model reads 
\begin{equation}
    J_\alpha = \frac{1}{\Delta x^2}\big({J}_0 + 2 {J}_1 \big)= \frac{1}{\Delta x}\Big(
    \lambda_0^2 {\tilde \lambda_0} \sum_k u_k^* \tilde{p}_{0_k} u_k  + 2 \lambda_0^2 {\tilde\lambda_{1}}  \sum_k u_k^* \tilde{p}_{1_k}({\Ab} \ub)_k \Big) \, . 
    \label{eq:NonconstantM_Jalpha}
\end{equation}

\subsection{Convective Kinematics}
\label{sec:qcfd:convection}
Convection is subject to velocity-induced transport ($y=v$) in \Eq{eq:WResidualEq}, i.e., 
\begin{equation}
\displaystyle{\int}_{\Omega} y \Big(y \, \frac{\partial y}{\partial x}\Big)dx\, , 
\label{eq:conexampl}
\end{equation} 
resulting in a non-symmetric discrete operator that have to be treated explicitly, cf. Sec.~\ref{sec:Math}.
To treat the convective term, a first-order \textsc{UDS} or higher-order upwind-biased schemes \cite{Ferziger2020, Waterson2007} are commonly used. The particular higher-order schemes considered here refer to \textsc{CDS}, \textsc{LUDS}, and \textsc{QUICK} scheme~\cite{Johnson1992}, where the \textsc{CDS} is of course not upwind-biased. 
Knowledge about the flow direction is mandatory to apply upwind-biased approximations. Classically, the \textsc{UDS} follows from a $\max$~operation as given for the discretized approximation of the weighted residual formulation in \Eq{eq:conexampl}, viz.
\begin{equation}
    J_\text{UDS} = \displaystyle{\sum}_k \Big(\max [y_k, 0] \,  \underbrace{
    y_k\big(y_{k} - y_{k-1}\big)}_{\Eq{eq:backward}} \; 
 - \max [-y_k, 0] \, \underbrace{
 y_k \big(y_{k+1} - y_{k}\big)}_{\Eq{eq:forward}} \Big) 
 \,.
\label{eq:UDS}
\end{equation}

In the quantum model, the $\max$~operation is substituted by the discrete mask functions $\mb^{+}, \mb^{-}$, whose entries are either $0$ or $1$ depending on the flow direction. Here, the superscript~$(+)$ indicates a positive and the superscript~$(-)$ a negative flow direction along the coordinate axis. Mind that the mask function entries are linearly dependent, i.e., $m_{k}^+=1-m_{k}^-$. The application of $\mb^{+}$ and $\mb^{-}$ to the velocity~$y_k$ is performed by a pointwise multiplication $\big(\max[y_k, 0] = m_{k}^+ y_k\big)$ and $\big(\max[-y_k, 0] = -m_{k}^- y_k\big)$, realized with the circuit outlined in \Fig{fig:qnpu:potential}. This is because the masking operator is generally non-unitary, except for unidirectional flows where either $\mb^{+}$ or $\mb^{-}$ resembles the identity matrix. The implementation based on the circuit depicted by \Fig{fig:qnpu:potential} bypasses the unitarization problem without additional efforts, such as block encoding~\cite{Camps2024}. This \textsc{QNPU} realizes a non-unitary point-wise multiplication, with $\mb^{+}$ and $\mb^{-}$ encoded as normalized states in the potential gate~$P^\dagger(\bar{\lamb}_c)$, cf. \Fig{fig:qnpu:potential_potential_adder}. This is equivalent to a multiplication with a non-unitary diagonal matrix~\cite{Lubasch2020}. The underbraced terms in \Eq{eq:UDS} are identical to the backward and forward differencing expressions discussed in Sec.~\ref{sec:qcfd:diffschemes}. 

Given an arbitrary one-dimensional periodic flow field, the weak formulation of the \textsc{UDS} as a matrix-vector multiplication reads equivalent to the inhomogeneous material treatment in Sec.~\ref{sec:qcfd:material} as 
 \begin{equation}
\begin{split}
 &\begin{bmatrix}
        y_1\\
        y_2\\
        \vdots\\
        y_{N_\text{p}}
    \end{bmatrix}^\intercal
     \begin{bmatrix}
         m^+_1\\
         & m_2^+ \\
         && \ddots \\
         &&& m_{N_\text{p}}^+
     \end{bmatrix}
     \begin{bmatrix}
         y_1\\
         & y_2 \\
         && \ddots \\
         &&& y_{N_\text{p}}
     \end{bmatrix} 
   \begin{bmatrix}
         1 &  0     &  \dots  &-1 \\
        -1 &  1     &  \dots  & 0 \\
           & \ddots &  \ddots &   \\
         0 & \dots  & -1      & 1 \\
    \end{bmatrix}
 \begin{bmatrix}
        y_1\\
        y_2\\
        \vdots\\
        y_{N_\text{p}}
    \end{bmatrix}\\
    &- \begin{bmatrix}
        y_1\\
        y_2\\
        \vdots\\
        y_{N_\text{p}}
    \end{bmatrix}^\intercal
    \begin{bmatrix}
         -m^-_1\\
         & -m_2^- \\
         && \ddots\\
         &&& -m_{N_\text{p}}^-
     \end{bmatrix} 
     \begin{bmatrix}
         y_1\\
         & y_2 \\
         && \ddots\\
         &&& y_{N_\text{p}}^{}
     \end{bmatrix}  
       \begin{bmatrix}
        -1 &      1 &  \dots &  0 \\
           &     -1 &      1 &    \\
           &        & \ddots &  \ddots \\
        1  &  \dots &  0     & -1 
    \end{bmatrix}  \begin{bmatrix}
        y_1\\
        y_2\\
        \vdots\\
        y_{N_\text{p}}
    \end{bmatrix}  \\     
     \allowdisplaybreaks 
      =&    \begin{bmatrix}
        y_1\\
        y_2\\
        \vdots\\
        y_{N_\text{p}}^{}
    \end{bmatrix}^\intercal
    \begin{bmatrix}
        -m_1^-y_1 + m_1^+y_1 & m_1^-y_1& \hdots &-m_1^+y_1 \\
        -m_2^+y_2 &-m_2^-y_2+m_2^+y_2 &m_2^-y_2 &\\
         &\ddots &\ddots&\ddots \\
        m_{N_\text{p}}^-y_{N_\text{p}}^{} &\hdots &-m_{N_\text{p}}^+y_{N_\text{p}}^{}&-m_{N_\text{p}}^-y_{N_\text{p}}^{} +m_{N_\text{p}}^+y_{N_\text{p}}^{}
    \end{bmatrix} 
    \begin{bmatrix}
        y_1\\
        y_2\\
        \vdots\\
        y_{N_\text{p}}
    \end{bmatrix} \, .
    \label{eq:UDS_matrix}
    \end{split}
 \end{equation}
The corresponding discrete weighted residual expression of the band matrix with non-constant diagonals in \Eq{eq:UDS_matrix} is 
\begin{equation}
    J_\text{UDS} =
    \Bigg(\sum_k y_k \underbrace{\left(-m_k^-+m_k^+\right)y_k^{}}_\text{diagonal} y_k- \sum_k y_k\underbrace{m_k^+y_k^{}}_{\substack{\text{sub-}\\\text{diagonal}}}y_{k-1}^{}  + \sum_k y_k\underbrace{m_k^-y_k^{}}_{\substack{\text{super-}\\\text{diagonal}}}y_{k+1}^{}\Bigg) \, , 
    \label{eq:J_UDS_single}
\end{equation}
where $\left(-m_k^-+m_k^+\right)y_k$ simplifies to $|y_k|$. The products of $y_k$ and $m_k^{+/-}$, which arise along the off-diagonals, require using the circuit introduced in \Fig{fig:qnpu:potential_potential_adder}, i.e., $a_4 = y_k \, m_k^{+/-}$. 

The method's trade-off is the previous knowledge of~$\yb$ to construct the mask parameters $\mb^{-},\mb^{+}$ that are required for the quantum circuits evaluation. For a passive scalar transport, the advection velocity is known a priori but in a general nonlinear convection problem, the extraction of the velocity directions is a challenging task for the \textsc{QC}. This can be done, for example, by solving an additional level-set differential equation \cite{Osher2001}. Alternatively, as in this study, one can combine state tomography \cite{Nielsen2010} with a training approach, analogous to the procedure described above for \Eq{eq:OptAnsatz}. To this end, the required quantum encoding of a previous or last known velocity again uses the encoding denoted in \Eq{eq:Generic_Ansatz}.

Higher-order methods are derived by extending the information to more remote neighboring points. For the sake of clarity, the following discussion is restricted to a velocity in the positive coordinate direction. The central location~$k$ in the finite approximation is referred to as $C$. The related upstream location is depicted by $U \leftarrow k-1$, while the downstream location refers to $D \leftarrow k+1$. To increase the order of the \textsc{FD} approximation, information from a second (remote) upstream location is required, i.e., $UU \leftarrow k-2$. The strategy introduced for the \textsc{UDS} can be extended by exchanging the approximation of the derivative ${\partial y}/{\partial x}|_C$ in \Eq{eq:UDS} with 
\begin{align}
    &\frac{y_C-y_U}{\Delta x} + \mathcal{O}(\Delta x)\,\label{eq:UDS_scheme} \hphantom{xxxxxxxxxxxxxxxxx}\; \,\to  \hphantom{xx}\text{\textsc{UDS}} \,,\\
    &\frac{y_D-y_U}{2\Delta x}+ \mathcal{O}(\Delta x^2)\, \label{eq:CDS_scheme}\hphantom{xxxxxxxxxxxxxxxxx} \to \hphantom{xx} \text{\textsc{CDS}}\,,\\
    &\frac{3y_C-4y_U+y_{UU}}{2\Delta x} + \mathcal{O}(\Delta x^2) \, \label{eq:LUDS_scheme} \hphantom{xxxxxxxx}\;\;\, \to \hphantom{x} \text{\textsc{LUDS}}\,, \\
    &\frac{2y_D+3y_C-6y_U+y_{UU}}{6\Delta x}+ \mathcal{O}(\Delta x^3) \, \label{eq:QUICK_scheme} \quad\;\;\; \to\, \text{\textsc{QUICK}} \, . 
\end{align}

Obviously, the matrix bandwidth~$b_\text{w}$ increases for the higher-order schemes in comparison to the \textsc{UDS} example in \Eq{eq:UDS_matrix} ($b_\text{w}=3$). While the symmetric \textsc{CDS} remains compact ($b_\text{w}=3$), the \textsc{LUDS} results in a bandwidth of $b_\text{w}=5$ and the \textsc{QUICK} \cite{Johnson1992} stencil recovers a larger band of $b_\text{w}=7$. The implementation applies the ideas presented in Sec.~\ref{sec:qcfd:diffschemes}. The shifted indices, e.g., $k-1$, are realized analog to Eqns.~(\ref{eq:forward_qc}-\ref{eq:backward_qc}) by using the shift operation. Similarly, a sequence of two adders is used to realize a $k\pm2$~shift. Again, we use a mask function as outlined in \Eq{eq:UDS}. The cost function contributions for the above-mentioned schemes for positive/negative velocities read 
\begin{equation}
\begin{split}
\label{eq:J_UDS}
    J_\text{UDS} &= \underbrace{\lambda_0^2 \tilde{\lambda}_0}_{\lambda_0^3} \sum_k u_k^* \underbrace{\tilde{p}_{k_0}}_{|u_k|} u_k- \lambda_0^2  \sum_k u_k^* \underbrace{\tilde{\lambda}_{-1}\tilde{p}_{{-1}_k}}_{m_k^+y_k^{}} (\Ab^\dagger \ub)_k + \lambda_0^2  \sum_k u_k^* \underbrace{\tilde{\lambda}_1\tilde{p}_{1_k}}_{m_k^-y_k^{}}(\Ab \ub)_{k}^{}\,, \hphantom{xxx}
\end{split}
\end{equation}
\begin{equation}
    \begin{split}
    2\, J_\text{CDS} &= -\lambda_0^2  \sum_k u_k^* \underbrace{\tilde{\lambda}_{-1}\tilde{p}_{{-1}_k}}_{m_k^+y_k^{}} (\Ab^\dagger \ub)_{k}^{} + \lambda_0^2  \sum_k u_k^* \underbrace{\tilde{\lambda}_1\tilde{p}_{1_k}}_{m_k^+y_k^{}} (\Ab \ub)_{k}^{}\hphantom{xxxxxxxxxxxxxxxxxxxxxxxx}\\
    &\hphantom{xxxxxxxxxxxxxxxxx}\,\quad+\lambda_0^2  \sum_k u_k^*  \underbrace{\tilde{\lambda}_{-1}\tilde{p}_{{-1}_k}}_{m_k^-y_k^{}} (\Ab^\dagger \ub)_{k}^{} - \lambda_0^2  \sum_k u_k^* \underbrace{\tilde{\lambda}_{1}\tilde{p}_{1_k}}_{m_k^-y_k^{}}(\Ab \ub)_{k}^{}\,,
    \label{eq:J_CDS}
    \end{split}
\end{equation}
\begin{equation}  
\begin{split}
\label{eq:J_LUDS}
    2\, J_\text{LUDS} &=   {3} \underbrace{\lambda_0^2 \tilde{\lambda}_0}_{\lambda_0^3} \sum_k u_k^* \underbrace{\tilde{p}_{0_k}}_{|u_k|} u_k \hphantom{xxxxxxxxxxxxxxxxxxxxxxxxxxxxxxxxxxxxxxxxxxxxxx}\\
    &\hphantom{xxxxxxxxxxx} - {4}\lambda_0^2  \sum_k u_k^* \underbrace{\tilde{\lambda}_{-1}\tilde{p}_{{-1}_k}}_{m_k^+y_k^{}} (\Ab^\dagger \ub)_{k} + \lambda_0^2  \sum_k u_k^*  \underbrace{\tilde{\lambda}_{-2} \tilde{p}_{{-2}_k}}_{m_k^+y_k^{}} (\Ab^{\dagger^2} \ub)_{k}\\
   &\hphantom{xxxxxxxxxxxxxxxxxxxxxxxxxxxx}\quad\;\, + {4}\lambda_0^2  \sum_k u_k^* \underbrace{\tilde{\lambda}_1\tilde{p}_{1_k}}_{m_k^-y_k^{}} (\Ab \ub)_{k}^{}  - \lambda_0^2  \sum_k u_k^* \underbrace{\tilde{\lambda}_2\tilde{p}_{2_k}}_{m_k^-y_k^{}}(\Ab^{^2}\ub)_{k}^{}\,,
 \end{split}
 \end{equation}
\begin{equation}
 \begin{split}
     6\, J_\text{QUICK} &= {3}\underbrace{\lambda_0^2 \tilde{\lambda}_0}_{\lambda_0^3} \sum_k u_k^* \underbrace{\tilde{p}_{0_k}}_{|u_k|} u_k \\
     &\quad + {2} \lambda_0^2  \sum_k u_k^* \underbrace{\tilde{\lambda}_1\tilde{p}_{1_k}}_{m_k^+y_k^{}}(\Ab \ub)_{k}^{}  - 6\lambda_0^2  \sum_k u_k^*   \underbrace{\tilde\lambda_{-1}\tilde{p}_{{-1}_k}}_{m_k^+y_k^{}} (\Ab^\dagger \ub)_{k}^{} 
     + \lambda_0^2  \sum_k u_k^*  \underbrace{\tilde{\lambda}_{-2} \tilde{p}_{{-2}_k}}_{m_k^+y_k^{}} (\Ab^{\dagger^2} \ub)_{k}^{} \\
     &\quad\qquad -   {2}\lambda_0^2  \sum_k u_k^*  \underbrace{\tilde{\lambda}_{-1} \tilde{p}_{{-1}_k}}_{m_k^-y_k^{}} (\Ab^\dagger \ub)_{k}^{} + 6\lambda_0^2  \sum_k u_k^* \underbrace{\tilde{\lambda}_1 \tilde{p}_{1_k}}_{m_k^-y_k^{}} (\Ab \ub)_{k}^{}
     - \lambda_0^2  \sum_k u_k^* \underbrace{\tilde{\lambda}_2 \tilde{p}_{2_k}}_{m_k^-y_k^{}}(\Ab^{^2}\ub)_{k}^{}\,.
     \label{eq:J_QUICK}
\end{split}
\end{equation}

In the case of a passive scalar transport, the main diagonal contribution becomes associated with the prescribed transport velocity~$v$ ($p_{0_k} \to |v_k|$) and the mask function product simplifies as $m_k^{+/-} |y_k| \to m_k^{+/-} |v_k|$, which is inherently known. It is worth mentioning that \textsc{CDS} is independent of the flow direction by definition. Therefore, a reformulation of the cost function to $2 \, J_\text{CDS} = \left( \sum_k y_ky_k^{} y_{k+1}^{} - \sum_k y_k y_k^{} y_{k-1}^{}\right)$, without mask functions, is preferred to reduce computational cost. 

The upwind-biased, higher-order approximation methods discussed above may be too diffusive and are usually not monotonicity-preserving. Monotonicity issues are usually addressed by additional Total Variation Diminishing~(\textsc{TVD}) or Normalized Variable Diminishing~(\textsc{NVD}) corrections~\cite{Leonard1988}, which limit the higher-order contributions to preserve monotonicity. However, such corrections involve further $\min/\max$~operations that are unfavorable in the \textsc{QC} framework and are therefore not discussed here. 
A commonly practiced, simpler heuristic approach is the use of a flux-blending procedure~(\textsc{FB}), which blends two different approximation schemes. The scheme usually combines a dispersive scheme (e.g., \textsc{CDS}) and a diffusive scheme (e.g., \textsc{UDS}) using a heuristic blending factor $\beta \in [0,1]$, for example 
 \begin{equation}
   J_{\text{FB}_\beta} = (1-\beta) J_\text{UDS} + \beta J_\text{CDS} \, . 
   \label{eq:FluxB}
\end{equation}
As regards the dispersive scheme, one might also opt for more advanced central schemes to improve the predictive performance, e.g., \citeauthor{Tam1993}~\cite{Tam1993}, where the dispersion effects are minimized at the expense of order reduction. An interesting question refers to the convergence of the optimizer when using a flux-blending approach and the advantages that can be achieved in reducing the depth of the bricklayer ansatz, cf. Sec.~\ref{sec:hybrid}. These issues will be addressed in the results section of this paper, cf. Sec.~\ref{sec:Results}.

%%%%%%%%%%%%%%%%%%%%%%%%%%%%%%%%%%%%%%%%%%%%%
% Complexity
%%%%%%%%%%%%%%%%%%%%%%%%%%%%%%%%%%%%%%%%%%%%%
\section{\textcolor{brown}{Complexity}}
\label{sec:qcfd:gatecomplexity}
A critical requirement for feasibility is defined by the decoherence time, i.e., the interval in which a qubit remains stable before losing its quantum state due to environmental effects~\cite{BlackKuhnWilliams2002}. This time interval is generally short for current \textsc{QC}s, suggesting to limit the number of required quantum operations~\cite{Cerezo2021} by using shallow quantum circuits. To analyze the gate scalability of the proposed method for arbitrary hardware, the number of gates is evaluated in terms of single-qubit (\textsc{R$_z$,R$_y$,U},\dots) and two-qubit (\textsc{CNOT},\dots) gates, for all presented \textsc{QCFD}~circuits of Sec.~\ref{sec:Quantum}. Mind that the gate complexity study assesses the ansatz and \textsc{QNPU}~gates, states like the mask functions $\mb^{+/-}$ for the matrix contributions in $\pb_q$ are assumed to be realized efficiently by an appropriate ansatz.

Figures~\ref{fig:ansatz_complexities} and~\ref{fig:Complexities} provide the complexities for the ansatz and the \textsc{QNPU}s, respectively. Figure~\ref{fig:ansatz_complexities} indicates a linear scaling of both the gates and the control parameters of the ansatz presented in Sec.~\ref{sec:hybrid} w.r.t. the number of qubits $n$~\cite{BravoPrieto2023}. In Fig.~\ref{fig:Complexities}, the building blocks of Sec.~\ref{sec:Quantum}, cf.~$P$ (\Fig{fig:qnpu:potential}) and $A$ (Fig.~\ref{fig:qnpu:adder}), are marked by $\triangle$ and $\square$, respectively, while the lines correspond to the matrix's diagonals, e.g., $q \in \{1,2\}$.
\begin{figure}[htbp]
    \centering
    \tikzsetnextfilename{AnsatzComplexity}
\begin{tikzpicture}
\begin{semilogyaxis}[
tick pos=left,
xmajorgrids,
xlabel = {$n$ [\  ]},
ylabel = {[\  ]},
xmin=2, xmax=13.5,
ymin=2,
ymax=1500,
ymajorgrids,
yminorgrids,
height = 0.38\textwidth,
legend pos=north west,
legend style={draw=black!15!black,legend cell align=left, fill opacity=0.8, draw opacity=1, text opacity=1},
legend columns=5, 
]

\addlegendimage{semithick,solid,mark={o},only marks}
\addlegendentry{\scriptsize{$d=1,\,$}}

\addlegendimage{semithick,solid,mark={diamond},only marks}
\addlegendentry{\scriptsize{$d=3,\,$}}

\addlegendimage{semithick,solid,mark={triangle},only marks}
\addlegendentry{\scriptsize{$d=5,\,$}}

\addlegendimage{semithick,solid,mark={square},only marks}
\addlegendentry{\scriptsize{$d=7,\,$}}

\addlegendimage{semithick,solid,mark={+},only marks}
\addlegendentry{\scriptsize{$d=9\,$}}

% DATA   %%%%%%%%%%%%%%%%%%%%%%%%

\addplot [semithick,solid,mark={o}]
table [col sep = comma , row sep=crcr]{%
2.0,  12 \\  
4.0,  32 \\  
6.0,  52 \\  
8.0,  72 \\
10.0, 92 \\
12.0, 112 \\
};

\addplot [semithick,solid,mark={diamond}]
table [col sep = comma , row sep=crcr]{%
2.0,  28 \\  
4.0,  80 \\  
6.0,  132 \\  
8.0,  184 \\
10.0, 236 \\
12.0, 288 \\
};

\addplot [semithick,solid,mark={triangle}]
table [col sep = comma , row sep=crcr]{%
2.0,  44 \\  
4.0,  128 \\  
6.0,  212 \\  
8.0,  296 \\
10.0, 380 \\
12.0, 464 \\
};

\addplot [semithick,solid,mark={square}]
table [col sep = comma , row sep=crcr]{%
2.0,  60 \\  
4.0,  176 \\  
6.0,  292 \\  
8.0,  408 \\
10.0, 524 \\
12.0, 640 \\
};

\addplot [semithick,solid,mark={+}]
table [col sep = comma , row sep=crcr]{%
2.0,  76 \\  
4.0,  224 \\  
6.0,  372 \\  
8.0,  520 \\
10.0, 668 \\
12.0, 816 \\
};

\addplot [semithick,dashed,mark={o},mark options=solid]
table [col sep = comma , row sep=crcr]{%
2.0,  4 \\  
4.0,  10 \\  
6.0,  16 \\  
8.0,  22 \\
10.0, 28 \\
12.0, 34 \\
};

\addplot [semithick,dashed,mark={diamond},mark options=solid]
table [col sep = comma , row sep=crcr]{%
2.0,  8 \\  
4.0,  22\\  
6.0,  36 \\  
8.0,  50 \\
10.0, 64 \\
12.0, 78 \\
};

\addplot [semithick,dashed,mark={triangle},mark options=solid]
table [col sep = comma , row sep=crcr]{%
2.0,  12 \\  
4.0,  34 \\  
6.0,  56 \\  
8.0,  78 \\
10.0, 100 \\
12.0, 122 \\
};

\addplot [semithick,dashed,mark={square},mark options=solid]
table [col sep = comma , row sep=crcr]{%
2.0,  16 \\  
4.0,  46 \\  
6.0,  76 \\  
8.0,  106 \\
10.0, 136 \\
12.0, 166 \\
};

\addplot [semithick,dashed,mark={+},mark options=solid]
table [col sep = comma , row sep=crcr]{%
2.0, 20 \\  
4.0,  58 \\  
6.0,  96 \\  
8.0,  134 \\
10.0, 172 \\
12.0, 210 \\
};

% \addplot [thin,gray]
% table [col sep = comma , row sep=crcr]{%
% 2.0,  10\\  
% 4.0,  22 \\  
% 6.0,  70\\  
% 8.0,  262 \\
% 10.0, 1030 \\
% 12.0, 4112 \\
% };
% \node[right,align=left] at (8.5,1000) {\textcolor{gray}{$\mathcal{O}(2^n)$}};

% \addplot [thin,gray]
% table [col sep = comma , row sep=crcr]{%
% 2.0,  10\\  
% 4.0,  22 \\  
% 6.0,  42\\  
% 8.0,  70 \\
% 10.0, 106 \\
% 12.0, 150 \\
% 14.0, 202 \\ %196 \\
% };
% \node[right,align=left] at (12.4,260) {\textcolor{gray}{$\mathcal{O}(n^2)$}};

\addplot [thin,gray]
table [col sep = comma , row sep=crcr]{%
2.0,  4\\  
4.0,  16 \\  
6.0,  64\\  
8.0,  256 \\
10.0, 1024 \\
12.0, 4096 \\
};
\node[right,align=left] at (8.5,1000) {\textcolor{gray}{$\mathcal{O}(2^n)$}};

\addplot [thin,gray]
table [col sep = comma , row sep=crcr]{%
2.0,  4\\  
4.0,  16 \\  
6.0,  36\\  
8.0,  64 \\
10.0, 100 \\
12.0, 144 \\
14.0, 196 \\
};
\node[right,align=left] at (12.4,260) {\textcolor{gray}{$\mathcal{O}(n^2)$}};

\end{semilogyaxis}
\end{tikzpicture}
    \caption{Gate (solid lines) and parameter (dashed lines) complexity of the employed hardware efficient ansatz proposed by \citeauthor{BravoPrieto2023}~\cite{BravoPrieto2023} for circuit depths $\mathbf{d\in[1,9]}$.}
    \label{fig:ansatz_complexities}
\end{figure}
\begin{figure}[htbp]
     \centering
     \tikzexternaldisable
\begin{tikzpicture}
\begin{semilogyaxis}[
tick pos=left,
xmajorgrids,
xlabel = {$n$ [\  ]},
ylabel = {number of gates  [\  ]},
xmin=2, xmax=13.5,
ymin=10,
ymax=1500,
ymajorgrids,
yminorgrids,
height = 0.38\textwidth,
]

\addplot [ only marks, semithick,black,mark=square]
table [col sep = comma , row sep=crcr]{%
2.0,  16 \\  
4.0,  92 \\  
6.0,  154 \\  
8.0,  216 \\
10.0, 278 \\
12.0, 340 \\
};
\node[right] at (12,340) {\scriptsize{$A$ \Fig{fig:qnpu:adder}}};

\addplot [ only marks, semithick,black, mark=triangle]
table [col sep = comma , row sep=crcr]{%
2.0,  30 \\  
4.0,  60 \\  
6.0,  90 \\  
8.0,  120 \\
10.0, 150 \\
12.0, 180 \\
};
\node[right,align=left] at (12,180) {\scriptsize{$P$ \Fig{fig:qnpu:potential}}};

\addplot [ semithick,loosely dashed]
table [col sep = comma , row sep=crcr]{%
2.0,  46\\  
4.0,  152 \\  
6.0,  244\\  
8.0,  336 \\
10.0, 428 \\
12.0, 520 \\
};
\node[right,align=left] at (12,500) {\scriptsize{$A_p$ \Fig{fig:qnpu:potential_adder}}};

\addplot [ semithick,black,dashed]
table [col sep = comma , row sep=crcr]{%
2.0,  62\\  
4.0,  244 \\  
6.0,  398\\  
8.0,  552 \\
10.0, 706 \\
12.0, 860 \\
};
\node[right,align=left] at (12,850) {\scriptsize{$A^2_p$ \Fig{fig:qnpu:potential_adder}}};

\addplot [ semithick,black,dash dot dot]
table [col sep = comma , row sep=crcr]{%
2.0,  76\\  
4.0,  212 \\  
6.0,  334\\  
8.0,  456 \\
10.0, 578 \\
12.0, 700 \\
};
\node[right,align=left] at (12,650) {\scriptsize{$A_{pp}$ \Fig{fig:qnpu:potential_potential_adder}}};

\addplot [ semithick,black,dash dot]
table [col sep = comma , row sep=crcr]{%
2.0,  92\\  
4.0,  304 \\  
6.0,  488\\  
8.0,  672 \\
10.0, 856 \\
12.0, 1040 \\
};
\node[right,align=left] at (12,1200) {\scriptsize{$A^2_{pp}$ \Fig{fig:qnpu:potential_potential_adder}}};

\addplot [thin,gray]
table [col sep = comma , row sep=crcr]{%
2.0,  16\\  
4.0,  64 \\  
6.0,  256\\  
8.0,  1024 \\
10.0, 4096 \\
12.0, 16384 \\
};
\node[right,align=left] at (6.5,1000) {\textcolor{gray}{$\mathcal{O}(2^n)$}};

\addplot [thin,gray]
table [col sep = comma , row sep=crcr]{%
2.0,  16\\  
4.0,  64 \\  
6.0,  144\\  
8.0,  256 \\
10.0, 400 \\
12.0, 576 \\
};
\node[right,align=left] at (8.5,220) {\textcolor{gray}{$\mathcal{O}(n^2)$}};

% \addplot [thin,red,dashed]
% table [col sep = comma , row sep=crcr]{%
% 2.0,  16\\  
% 4.0,  72 \\  
% 6.0,  224\\  
% 8.0,  520 \\
% 10, 1008\\
% 12, 1736\\
% };

% \addplot [thin,red,dotted]
% table [col sep = comma , row sep=crcr]{%
% 2.0,  16\\  
% 4.0,  256 \\  
% 6.0,  4096\\  
% };

% \addplot [thin,gray]
% table [col sep = comma , row sep=crcr]{%
% 2.0,  60\\  
% 4.0,  120 \\  
% 6.0,  180\\  
% 8.0,  240 \\
% 10.0, 300 \\
% 12.0, 360 \\
% };
% \node[right,align=left] at (8.5,60) {\textcolor{gray}{$\mathcal{O}(n)$}};

\end{semilogyaxis}
\end{tikzpicture}
\tikzexternalenable
     \caption{Gate complexity of the employed \textsc{QNPU} assemblies in Sec.~\ref{sec:qnpus}.}
     \label{fig:Complexities}
\end{figure}
The building blocks of the \textsc{QNPU} are independent of the derivative matrix structure, allowing to reproduce any band matrix with excellent gate complexities, as it is shown by the evolution of the symbols in \Fig{fig:Complexities}. In particular, the main diagonal complexity is based on the \textit{polylog} scaling, $\mathcal{O}(log(n)^r) : r \in \mathbb{R}$, of the potential circuit~$P$. Similarly, the off-diagonals benefit from the same \textit{polylog} complexity of the adder circuit~$A$ and the potential~$P$. It is also evident that the scaling is independent of the bandwidth~$b_\text{w}$, which encourages the application of higher-order approximation schemes.

The proposed algorithm implements generic band matrices with linear ansatz scaling and \textit{polylog} complexity for the compiled circuits of Sec.~\ref{sec:qnpus}, which are both more efficient than a quadratic ($n^2$) scaling. This is clearly illustrated by the slope of the gray $\mathcal{O}(n^2)$ reference line in Fig.~\ref{fig:ansatz_complexities} and Fig.~\ref{fig:Complexities}, making these quantum circuits tailored for \textsc{VQA} implementations. 
The method therefore favorably distinguishes from other variational methods based upon Linear Combination of Unitaries (\textsc{LCU}) \cite{Crowe2022}, where the number of ancilla qubits and measurements scales with the \textsc{LCU} contributions.
Moreover, the proposed approach allows the optimization of the gate sequences for a specific hardware, thus further improving the complexity. To guarantee this scaling w.r.t. the number of qubits, the efficient amplitude encoding by a suitable ansatz is essential \cite{Chen2021,Kyriienko2021,Mozafari2022,Creevey2023,Melnikov2023,Araujo2021}.

To estimate the \textsc{VQA} time complexity $\mathcal{T}$, the individual contributions for state preparation $\mathcal{T}_p$, cost function $\mathcal{T}_c$, derivative approximation $\mathcal{T}_d$, and optimization $\mathcal{T}_o$ are applied \cite{Sato2021} while measurement complexity is neglected due to the noise-free state vector simulation. 
The time complexity for state preparation scales with the total circuit depth as $\mathcal{T}_p\propto\mathcal{O}(d + log(n)^r/n)$, with $d$ for the ansatz's depth and the gates per qubit ratio for the \textsc{QNPU}. 
The number of cost function contributions depends on the number of differential operations~$d_\text{o}$ in \Eq{eq:GoverningEq} and the bandwidth~$b_\text{w}$ of the employed differencing stencil, which results in a scaling independent of $n$, $\mathcal{T}_c \propto \mathcal{O}(d_\text{o} b_\text{w})$.
For approximating the cost function's derivative, additional evaluations proportional to the number of parameters $c \propto \mathcal{O}(d n)$ are required, such that $\mathcal{T}_d \propto \mathcal{O}(c) \mathcal{T}_c = \mathcal{O}(c d_\text{o} b_\text{w})=\mathcal{O}(d n d_\text{o} b_\text{w})$. 
In the iterative optimization, the above operations are executed in every step, with a complexity $\mathcal{T}_o$ defined by the underlying method and the initial guess. The total time complexity results from the product of the individual contributions~\cite{Sato2021} in $\mathcal{T} = \mathcal{O}\Big((d + log(n)^r/n) \big(d_\text{o} b_\text{w} (1+dn)\big)  \mathcal{T}_o\Big)$, where $(\mathcal{T}_c+\mathcal{T}_d)=\big(d_\text{o} b_\text{w} (1+dn)\big)$ accounts for the total number of circuits required for gradient-based optimizations. Note that $\mathcal{T}$ is the time complexity per time step $l$ while the overall method complexity scales linearly with $N_\text{t}$, matching with Refs.~\cite{Sato2021,Leong2022}.

Unlike fault-tolerant \textsc{QC}s \cite{Brearley2024,Hu2024,Over2024b,Sato2024,Wright2024}, where the circuit depth scales linearly with the number of time steps~$N_\text{t}$, the \textsc{VQA} scaling is time-independent due to its hybrid classical-quantum nature. For explicit \textsc{CFD} applications on \textsc{QC}s, time complexity is particularly critical since local spatial refinements inherently yield smaller time steps. Another major benefit of \textsc{VQA}s is the ability to efficiently model nonlinearities, circumventing the no-cloning theorem by creating multiple copies of the state through parameterization using ans\"{a}tze \cite{Lubasch2020}. In contrast, fault-tolerant approaches are limited to linear problems otherwise the number of gates scales exponentially with time ~\cite{Esmaeilifar2024}. A key feature of \textsc{VQA} is the use of variational parameters to mitigate I/O problems of \textsc{QC}, where fault-tolerant approaches require efficient state preparation and tomography, potentially destroying quantum advantage. For example, reproducing the solution on a quantum computer -- required for each shot of each measurement --  the whole simulation must be repeated in case of fault-tolerant frameworks. In contrast, the \textsc{VQA} approach allows efficient reconstruction of the solution directly from the variational parameters without repeating the simulation.

%###########################################
%       Numerical Results
%###########################################
\section{Numerical Results}
\label{sec:Results}
The section verifies \textsc{VQA} results against results obtained from a classical \textsc{FD} approach for different one-dimensional problems, i.e., heat transfer, nonlinear convective transport, Burgers' equation, and wave equation. The first two subsections briefly describe the overall \textsc{VQA} approach and the optimization procedure. Initial results are dedicated to heat transfer simulations using the example of transient heat conduction with inhomogeneous diffusivity (Sec.~\ref{sec:results:material}) and steady advective/conductive heat transfer at three different \textit{P\'{e}clet} numbers, i.e., $Pe_{\Delta x} = 0.3,\,3\text{, and } 30$ (Sec.~\ref{sec:results:steady_diff_conv}). Subsequently, the nonlinear momentum transfer is studied for uni- and bidirectional inviscid cases (Sec.~\ref{sec:Results:Riemann}/~\ref{sec:Results:saw}), and a viscous case at $Re=100$ (Sec.~\ref{sec:Results:Burgers}). Finally, Section~\ref{sec:results:wave} reports results for the one-dimensional wave equation. Depending on the properties of the cost function contribution in \Eq{eq:WResidualEq}, either implicit or explicit approaches are employed, as discussed in Sec.~\ref{sec:Math}. The source term contribution~$f$ in \Eq{eq:GoverningEq} is not the subject of this work and is therefore not considered in the applications. An extension to include this contribution is possible, as shown in Ref.~\cite{Over2024}.

Presented results employ $N_\text{p}=16$ and $N_\text{p}=64$ interior points using $n=4$ and $n=6$ qubit registers, respectively. The fine discretization employs a total of $23$~qubits for the full circuit, including six register qubits, a copy of the register to handle the nonlinearity, six qubits for the mask function, and five auxiliary qubits, i.e., one ancilla for the Hadamard test and four carry qubits for the adder network. On a classical emulator, this requires modeling of linear algebra operators with sizes of $2^{23} \times 2^{23}$ and approaches the simulation limits on classical hardware. The study is based on the previously described \textsc{FD} approximations, which were used both for the classical simulation and for the development of the analogue \textsc{QNPU}s employed by the \textsc{VQA}. Deviations between the \textsc{VQA} result and the classical result are assessed using the l$_2$-error and the trace distance error
\begin{equation}
    \varepsilon_{\text{l}_2} = \sqrt{\sum_k\big(y_k^\text{FD}-y_k^\text{VQA}\big)^2}\, , \quad
   \varepsilon_\text{tr}=\sqrt{\Bigg(1- \Bigg|\sum_k\frac{y_k^\text{FD}}{\|\yb^\text{FD}\|_2} \frac{y_k^\text{VQA}}{\|\yb^\text{VQA}\|_2}\Bigg|^2 \Bigg)} \, .
   \label{eq:Results:errors}
\end{equation}
Error measures given without further indication only sum over spatial locations, as in \Eq{eq:Results:errors}, while additionally time-averaged expressions are denoted by $\bar\varepsilon_{\text{l}_2}$ or $\bar\varepsilon_\text{tr}$.
Trace-distance error levels found in the literature show values up to $\varepsilon_\text{tr} \sim 10^{-2}$ for standard \textsc{FD} approximations of similar one-dimensional problems \cite{Sato2021,Leong2022,Over2024}. An important aspect of this study is therefore (a) the associated quality of the predictions and (b) the dependence of the error level on resolution, that is, $n=4$ vs. $n=6$.

\subsection{\textsc{VQA} Approach}
The employed parameterization $U(\lamb_c^l)$ of the quantum state {$\ket{u^l} = U(\lamb_c^l) \ket{0}$} is restricted to the hardware-efficient fixed-structure ansatz of Ref.~\cite{BravoPrieto2023}. The latter is outlined in Sec.~\ref{sec:hybrid} and assembled in a brick-layer structure of depth $d\in[3,9]$. Using a real-valued ansatz for real-valued solutions, facilitates a lower number of trainable parameters and reduces the optimization complexity. 
Note that the ansatz is fully modular and can of course be replaced. Since the focus is on the developed \textsc{QNPU}s, the ansatz is not analyzed in detail here. In general, lower-depth circuits are preferred to limit computational effort and to enable near-term real hardware implementations. However, higher resolutions and error levels might motivate to increase the depth of the ansatz and simultaneously reduce the amount of qubits to lower the computational resources. Mind that enhancing the depth increases the expressiveness which might be required when increasing the resolution, cf. Sec. \ref{sec:optistrat}.

We would like to point out again that we assume a periodic pattern of all matrices as a reference starting point for the \textsc{QNPU} developments. In this regard, the implementation of non-periodic boundary conditions takes place in two steps \cite{Over2024}. First, inherently periodic contributions of the \textsc{QNPU}s along the boundaries are neutralized by a correction of the objective functional called $J_{\text{DN}}$. Subsequently, the objective functional is modified a second time to take into account the respective Dirichlet or Neumann contributions. Regarding the \textsc{QNPU}s for treating convective kinematics, boundary conditions could also be enforced by manipulating the mask function. In line with Ref.~\cite{Over2024}, transient influences are represented by cost function contributions to potential~$J_\text{P}$ and source~$J_\text{S}$ terms. 

The measurement of the quantum states neglects noise and simplifies the procedure to the trace-out of the ancilla qubit~\cite{Nielsen2010}. The corresponding density matrix $\rho_\text{ancilla}$ is recovered by a single shot of the state vector simulation.

\subsection{Optimization Strategy}
\label{sec:optistrat}
The parameters $\big(\lambda_0^l, \lamb_c^l\big)^\intercal$ are optimized in a global-to-local procedure, sequentially applying a zero-order Particle-Swarm Optimization (\textsc{PSO}) \cite{Kennedy1995} followed by a quasi-Newton method \cite{Nocedal2006}. In this two-step strategy, the global search of the gradient-free initial optimizer provides a trained starting point for the subsequent gradient-based local optimization. The preceding global optimization may avoid potential problems of local optimizers due to vanishing gradients, also called Barren Plateaus (\textsc{BP}s). The global optimizer is initialized with approximately $100$ samples randomly distributed over the optimization manifold. These samples are updated for $10$ iterations without considering gradient information and the most optimal set is used to initiate the local optimization. For subsequent time steps, the optimal parameters ($\upb^{l-1}$) of the previous time step serve as initialization for successive restarts, and the \textsc{PSO} step is skipped. The gradient-based (local) optimization employs a version of the well-established \textsc{BFGS} approach~\cite{Nocedal2006}. The convergence of the gradient-based optimizer is judged by the l$_2$-norm of the derivative concerning a defined tolerance, i.e., $\texttt{tol} \, > \|\nabla J(\upb^l)\|_{\text{l}_2}$.

The derivative of the cost function is approximated by the parameter-shift rule \cite{Schuld2019,Banchi2021,Crooks2019}. This method has the advantage of reusing the existing quantum circuits, allowing the derivative to be computed by evaluating the cost function $2c$~times, with each parameter perturbed by ${\pm{\pi}/2}$. Consequently, the computational cost of the derivative scales with the number of control variables~$c$, which in turn scales with the number of qubits $n$ and the ansatz depth~$d$ as $c\propto\mathcal{O}(nd)$, cf. Fig.~\ref{fig:ansatz_complexities}. 

Ideally, the parameterized quantum circuits used here as an approach for solving variational problems should be highly expressive, allowing a good approximation to the desired solution. This is typically achieved by increasing the depth $d$ of the approach with increasing resolution. On the other hand, the approach must have sufficiently large gradients to enable training and avoid stagnating in \textsc{BP}s, which pose a well-known challenge to optimization in \textsc{VQA}s due to their association with exponentially vanishing gradients \cite{Larocca2024,Pool2024}. A balance must be struck between expressiveness and trainability. In general, variational circuits are not unique, which permits to design their configuration to partially mitigate the onset of \textsc{BP}s. Nonetheless, it is well known that highly expressive ansätze necessarily lead to \textsc{BP}s \cite{Holmes22}. While the results of this study do not show severe accuracy problems, resolution influences on the trainability of the approaches are observed in the included examples. Thus, studies on larger systems will be required to deviate from the generic structure of a real-valued ansatz circuit used in this work. Options for improvements include, in addition to incorporating regularizers, the use of classical shadows \cite{Sack22}, using classical neural networks \cite{Friedrich22}, adding entangled ancilla qubits \cite{Yao25}, and targeting certain tensor network quantum states \cite{Barthel25}.

\subsection{Transient Heat Conduction with Non-Uniform Diffusivity}
\label{sec:results:material}
The first application refers to the transient heat conduction ($a_1=0, a_2=1$). It observes the evolution of the dimensionless temperature~$y=\vartheta(x,t)$ for an inhomogeneous material with non-constant thermal diffusivity~$a_3=\alpha(x)$, neglecting convective heat transport and source terms ($a_4=a_5=f=0$). Dirichlet boundary conditions are employed at both ends of the domain, with $\vartheta\big|_{x=0} = 0$ and $\vartheta\big|_{x=1}=1$, and the temperature is initialized with $\vartheta(x,t=0)=0$. The \textsc{VQA} simulations are conducted with a circuit depth of $d=5$, using $n=4$~qubits, and $d=9$ for $n=6$~qubits. The convergence criterion of the optimizer is set to $\texttt{tol} \,= 10^{-3}$, limited to a maximum number of $400$ iterations per time step. 

The thermal diffusivity~$\alpha(x)$ follows a Gaussian distribution and reaches a maximum in the center of the domain, viz. $\alpha(x)=1+\exp \big(-100 \, (0.5-x)^2\big)$. The spatially varying heat flux is captured by the contributions ${p}_{0_k} = -\alpha_{k-1}-\alpha_{k}$ and $p_{1_k} = \alpha_k$ as indicated in \Eq{eq:diff_noconst_matrix}. Combining the cost function contributions of Sec.~\ref{sec:QCFD} with the contributions of the boundary conditions to $J_\text{P},\,J_\text{S}$ and the boundary specific contribution $J_{\text{DN}}$ as introduced in Ref.~\cite{Over2024}, the optimization problem to be solved per time step reads 
\begin{equation}
    \min\limits_{\upb} J(\upb),  \quad \text{where } \quad 
    J(\upb)  =\overbrace{-{J}_0 - 2 {J}_1}^{-J_\alpha \eqref{eq:NonconstantM_Jalpha}} + \underbrace{J_{\text{DN}_{\alpha}}+J_\text{S}}_{\text{b.c. influenced \cite{Over2024}}} +J_\text{P}\, .
\end{equation}
As mentioned above, the two-step implementation of the boundary conditions \cite{Over2024}, initially neutralizes the inherently periodic contributions of the \textsc{QNPU}s along the boundaries by $J_{\text{DN}_{\alpha}}$  and subsequently introduces contributions of the inhomogeneous Dirichlet condition at $x=1$ to the source term contribution~$J_\text{S}$.

\begin{figure}[htbp]
    \centering
    \begin{subfigure}[t]{0.475\textwidth}
        \includegraphics{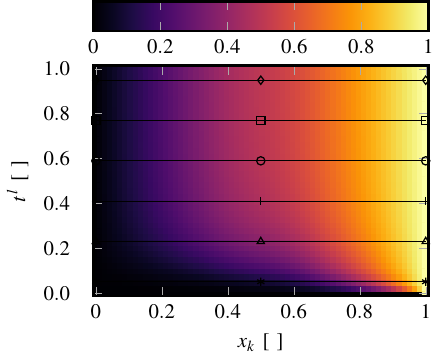}
        \caption{\textsc{VQA} predicted temperature evolution for $\pmb{N_\textrm{p}=64}$ with lines and symbols marking the equidistant time instants used to extract spatial temperature profiles in Subfig.~(\subref{fig:Results:Diffusivity:profiles}).}
        \label{fig:Results:Diffusivity:results}
    \end{subfigure}
        \hfill
    \begin{subfigure}[t]{0.475\textwidth}
        \centering
        \includegraphics{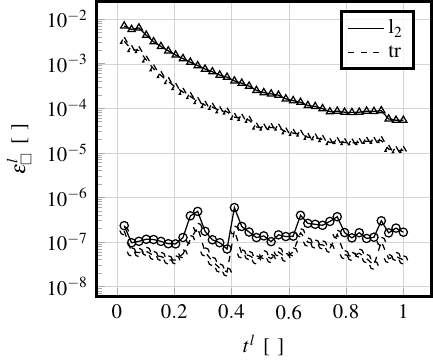}
        \caption{Temporal evolution of the error measures $\pmb{\varepsilon_{\text{l}_2}}$ and $\pmb{\varepsilon_\text{tr}}$ w.r.t. classical \textsc{FD} for $\pmb{n=4}$ ($\pmb{\circ}$) and $\pmb{n=6}$ ($\pmb{\triangle}$) qubits.}
        \label{fig:Results:Diffusivity:errors}
    \end{subfigure}
    \vskip\baselineskip
    \begin{subfigure}[t]{0.95\textwidth}
        \includegraphics{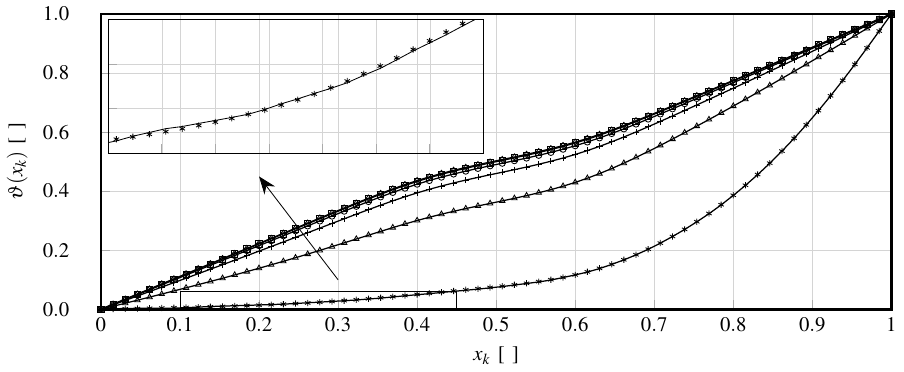}
        \caption{Spatial temperature profiles computed by the \textsc{VQA} (lines) and a \textsc{FD} method (symbols) for $\pmb{N_\textrm{p}=64}$ at the six time instants given in Subfig.~(\subref{fig:Results:Diffusivity:results}).}
        \label{fig:Results:Diffusivity:profiles}
    \end{subfigure}
    \caption{Comparison of \textsc{VQA} and \textsc{FD} results for the transient heat conduction problem with non-constant diffusivity. Results refer to $\pmb{N_\text{p}=64}$ interior grid points, using Dirichlet boundary conditions $\pmb{\big(\vartheta\big|_0 = 0,\, \vartheta\big|_1=1\big)}$ and zero initial temperature. The temporal evolution of the error measures in Subfig.~(\subref{fig:Results:Diffusivity:errors}) includes results for $\pmb{N_\text{p}=16}$ ($\pmb{\circ}$) and $\pmb{N_\text{p}=64}$ ($\pmb{\triangle}$) interior grid points.}
    \label{fig:Results:Diffusivity}
\end{figure}

Figure~\ref{fig:Results:Diffusivity} illustrates the results obtained for the \textsc{VQA} and the classical \textsc{FD} approach maintaining a constant time step of $\Delta t=0.018$ independent of the discretization. The spatio/temporal evolution of the \textsc{VQA}-predicted temperature~$\vartheta(x,t)$ for $N_\text{t}+1=40$ time steps and $N_\text{p} = 64$ is displayed in \Fig{fig:Results:Diffusivity:results}, where the abscissa refers to the spatial location and the ordinate to the normalized time. The figure also compares the spatial temperature profiles obtained from the \textsc{VQA} for $n=6$ with the corresponding results of the classical method in \Fig{fig:Results:Diffusivity:profiles} for six exemplary time instants. Figure~\ref{fig:Results:Diffusivity:errors} shows the evolution of the two error measures introduced in \Eq{eq:Results:errors} for $N_\text{p}=16$ ($n=4$) and $N_\text{p}=64$ ($n=6$) interior grid points along a normalized time horizon. An excellent visual agreement between \textsc{VQA} and \textsc{FD} methods is indicated by \Fig{fig:Results:Diffusivity:profiles}, which demonstrates that the proposed quantum model for non-homogeneous material properties can accurately reproduce classical simulation results. The inclination of the steady state profiles is halved in the center of the domain due to the doubling of the diffusivity. For $n=4$ qubits we obtain a remarkable agreement, as shown by the error levels presented in Fig.~\ref{fig:Results:Diffusivity:errors}. Nevertheless, the error levels determined for the finer resolution significantly exceed those of the coarser resolution. As the system approaches a steady state, both the l$_2$-error and the trace distance decrease to values below $10^{-6}$ and $10^{-4}$ for $n=4$ and $n=6$ qubits, respectively, cf.~\Fig{fig:Results:Diffusivity:errors}. The temporal averages read $\bar\varepsilon_{\text{l}_2} = 1.91 \times 10^{-7}$ and $\bar\varepsilon_\text{tr} = 7.04 \times 10^{-8}$ for $n=4$ qubits, while we obtain $\bar\varepsilon_{\text{l}_2} = 1.1 \times 10^{-3}$ and $\bar\varepsilon_\text{tr} = 3.1 \times 10^{-4}$ for $n=6$ qubits. The optimization appears to be more challenging for $n=6$ than for $n=4$ where more pronounced residuals are still detectable for $n=6$. These are noticeable as small deviations close to the left boundary during the initial time steps, cf. detail in \Fig{fig:Results:Diffusivity:profiles}. Notably, the solution in this area is not very demanding, and fluctuations around the almost constant value can be seen. Due to the successive improvement of the optimization during subsequent time steps and the steeper solution profile, these deviations become smaller as the steady state is approached.

\subsection{Steady Advection-Diffusion}
\label{sec:results:steady_diff_conv}
The second case covers the steady ($a_1=a_2=0$), source-free ($f=a_5=0$) advection-diffusion problem. The application describes the spatial temperature distribution~$y=\vartheta(x)$ of a fluid that is advected with a constant positive velocity~$v=a_4$. Diffusive transport is based on a constant thermal diffusivity~$\alpha=a_3=1$, and the problem is subject to Dirichlet conditions at both ends of the domain, viz. $\vartheta\big|_{x=0} = \vartheta_0 = 0$ and $\vartheta\big|_{x=1} = \vartheta_{N_\text{p}} = 1$. In this case, \Eq{eq:GoverningEq} reduces to an \textsc{ODE} with the analytical solution~\cite{Ferziger2020}
\begin{equation}
    \vartheta^\text{ANA}(x)= \vartheta_0 + \frac{e^{Pe_{\Delta x} \,/(\Delta x) \, x}-1}{e^{Pe_{\Delta x}L/\Delta x}-1} (\vartheta_{N_\text{p}} - \vartheta_0) \, , 
    \label{eq:analytical}
\end{equation}
where the local \textit{P\'{e}clet} number {$Pe_{\Delta x}=\Delta x \, v/\alpha$} describes the ratio between advective and diffusive transport. Hence, the \textsc{VQA} implementation can be verified against the classical approach and also validated against the analytical solution~\eqref{eq:analytical}. Mind that all numerical values in the tables compare \textsc{FD} with \textsc{VQA} data, while the line plots include the analytical solution and thereby allow an assessment of the physical correctness of the \textsc{FD}/\textsc{VQA} results.

The problem is discretized with $N_\text{p}=16$ interior grid points with $n=4$ qubits and a circuit depth of $d = 4$. The steady-state problem is iterated to convergence, achieved within {$\mathcal{O}$(10)} pseudotime iterations depending on $Pe_{\Delta x}$ and maintaining a reduction of the pseudotime residual by {$10^{-6}$}. The convergence criterion of the optimization in each pseudotime step is set to $\texttt{tol} \,= 10^{-6}$, limited to a maximum number of $200$ iterations. For a prescribed constant velocity~$v$, the advective term in \Eq{eq:GoverningEq} linearizes, such that the main diagonal contribution $p_{0_k}$ simplifies to $|v_k|$ and the off-diagonal contributions read $m_k^{+/-} |v_k|$, cf. Sec.~\ref{sec:qcfd:convection}. As in the previous case, the cost function employs contributions from the boundary condition by means of $J_{\text{DN}}$ and $J_\text{S}$~\cite{Over2024} in addition to
the diffusion $J_\text{cent-2}$ and the (explicit) contributions for the advective kinematics~$J_C$, viz. 
\begin{equation}
\min\limits_{\upb} J(\upb), \quad \text{where } \quad
J(\upb)  =-a_3 \overbrace{J_\text{cent-2}}^{\eqref{eq:seconddiff_quantum}}  
+a_4 \overbrace{J_C}^{(\ref{eq:J_UDS}-\ref{eq:J_QUICK})}
+ J_\text{P} 
+\underbrace{a_3 J_{\text{DN}_\text{cent-2}}
-a_4 J_{\text{DN}_C} + J_\text{S}}_{\text{b.c. influenced \cite{Over2024}}}
   \, .
\end{equation}

The advective contributions~$J_C$ and $J_{\text{DN}_C}$ take $C=\{\text{UDS},\,\text{CDS},\,\text{LUDS},\,\text{QUICK}, \, \text{FB}\}$ and adjust to the chosen advection scheme as introduced in Eqns.~(\ref{eq:J_UDS}-\ref{eq:FluxB}). The correction of the symmetric 2nd-order central difference approximation at the boundaries $J_{\text{DN}_\text{cent-2}}$ agrees for the left and right end of the domain.
The correction, labeled $J_{\text{DN}_C}$, of the periodic convective contribution~$J_C$ is implemented similarly. 
Again, the inhomogeneous Dirichlet condition at $x=1$ induces a contribution to the source term contribution~$J_\text{S}$. The solution is iterated to convergence by means of iterating the $J_\text{P}$ and $J_\text{S}$ contributions.

\begin{figure}[htbp]
    \centering
    \includegraphics{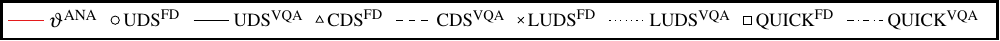}
    \vspace{1ex}
    
    \begin{subfigure}[t]{0.49\textwidth}
       \includegraphics{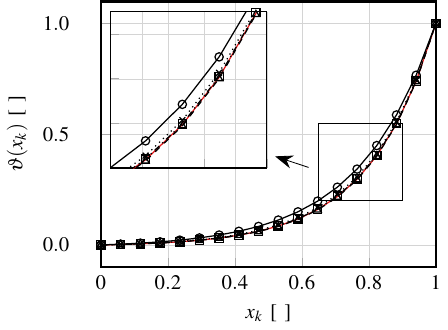}
        \caption{Solutions for $\pmb{Pe_{\Delta x} = 0.3}$.}
        \label{fig:Results:DiffConv:Pe03}          
    \end{subfigure}
    \hfill
    \begin{subfigure}[t]{0.49\textwidth}
        \centering
        \includegraphics{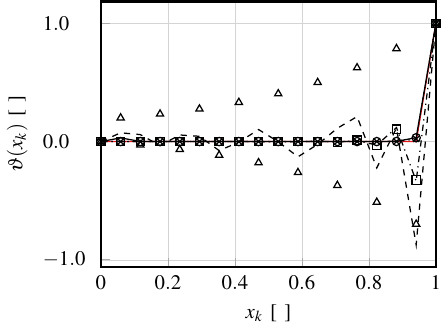}
        \caption{Solutions for $\pmb{Pe_{\Delta x}= 30}$.}
        \label{fig:Results:DiffConv:Pe30}
    \end{subfigure}
    \vskip\baselineskip
    \begin{subfigure}[t]{0.7\textwidth}
        \centering
        \includegraphics{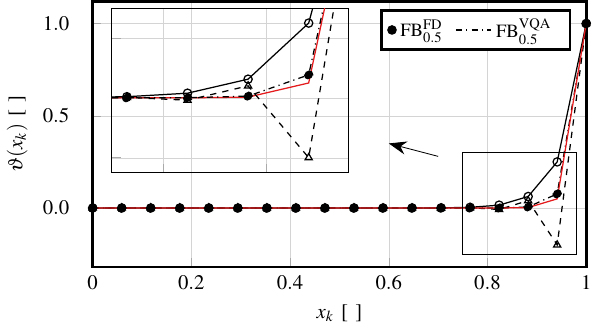} 
        \caption{Solutions for $\pmb{Pe_{\Delta x} = 3}$ using \textsc{FB} schemes with $\pmb{\beta \in \{0,0.5,1\}}$.}
        \label{fig:Results:DiffConv:Pe3}
    \end{subfigure}
    \caption{Results for the steady advection-diffusion problem obtained from the \textsc{VQA} (lines, $\pmb{d=4}$) and the \textsc{FD} (symbols) methods in comparison to the analytical solution (red-line) \pmb{\eqref{eq:analytical}} on a grid with $\pmb{N_\text{p}=16}$ interior points for three different local \textit{P\'{e}clet} numbers $\pmb{Pe_{\Delta x}}$.}
    \label{fig:Results:DiffConv}
\end{figure}

The test case serves to assess the performance of the convective schemes presented in Sec.~\ref{sec:qcfd:convection} w.r.t. the corresponding classical \textsc{FD}-implementation. For this purpose, simulations are conducted for the four basic schemes (\textsc{UDS}, \textsc{CDS}, \textsc{LUDS}, and \textsc{QUICK}) with two different \textit{P\'{e}clet} numbers, i.e., $Pe_{\Delta x}=0.3$ and $Pe_{\Delta x}=30$. Additional comparisons of \textsc{CDS}, \textsc{UDS}, and \textsc{FB} (with $\beta=0.5$), as introduced in Sec.~\ref{sec:qcfd:convection}, are computed for $Pe_{\Delta x}=3$. Figure~\ref{fig:Results:DiffConv} and Tab.~\ref{tab:Results:DIffConv:Errors} display the predictive performance of the \textsc{VQA} implementations in comparison to the results obtained by the classical method and analytical solutions, the latter is indicated by a red line. 

When attention is directed to the diffusion-dominated smaller \textit{P\'{e}clet} number as in \Fig{fig:Results:DiffConv:Pe03}, all convection schemes deliver an accurate description of the physics, with minor exceptions for the \textsc{UDS}, which displays a noticeable degree of numerical diffusion. As expected, the higher-order \textsc{CDS}, \textsc{LUDS}, and \textsc{QUICK} convection methods yield very good agreement with the analytical solution. For all investigated advective approximations, the predictive agreement of the \textsc{VQA} results and the classical results is excellent, which is also seen by the values in Tab.~\ref{tab:Results:DIffConv:Errors}. Supplementary to the l$_2$-results in Tab.~\ref{tab:Results:DIffConv:Errors}, the trace-distance results indicate sufficient expressibility of the employed ansatz. 

For the advection-dominated larger \textit{P\'{e}clet} number, the influence of the advective approximation becomes more pronounced, as given in \Fig{fig:Results:DiffConv:Pe30}. As expected, results of the unbiased \textsc{CDS} reveal instabilities due to lack of dominance of the main diagonal and the sign change of the secondary diagonals given by the stencil. The instabilities are seen for both classical and \textsc{VQA} results, though the agreement between them is naturally less exact, cf. Tab.~\ref{tab:Results:DIffConv:Errors}. The discrepancy between the classical and the \textsc{VQA} is smaller for the three upwind-biased schemes in comparison to \textsc{CDS}, which is also reflected in the $\varepsilon_{\text{l}_2}$ and $\varepsilon_{\text{tr}}$ values. The reproduction of the spurious peak-to-peak oscillations of the \textsc{FD}-based \textsc{CDS} requires a highly expressive ansatz in the \textsc{VQA} framework. To achieve a comparable agreement with other schemes, it demands larger ansatz depths, which leads to more challenging optimization problems. The results for \textsc{CDS} and \textsc{QUICK} scheme displayed in \Fig{fig:Results:DiffConv:Pe30} suffer from issues at the right end, whereas the schemes that avoid downwind influences, i.e., \textsc{LUDS} and \textsc{UDS}, provide more accurate results in the example chosen. Note that potential monotonicity issues of the \textsc{QUICK} and \textsc{LUDS}, responsible for generation of local extrema, are triggered here only for the \textsc{QUICK}, as confirmed by the good performance of \textsc{LUDS}.

The \textsc{FB} approach can balance the advantages and disadvantages of \textsc{CDS} and \textsc{UDS}, as shown in \Fig{fig:Results:DiffConv:Pe3}. The deviation of the \textsc{FD} from the \textsc{VQA} results is outlined in Tab.~\ref{tab:Results:DIffConv:Errors}. The differences observed between \textsc{UDS} and \textsc{CDS} results with increasing \textit{P\'{e}clet} number are also confirmed in the results of \Fig{fig:Results:DiffConv:Pe3}. Despite similar accuracy for $\varepsilon_{\text{l$_2$}}$, see Tab.~\ref{tab:OptPe3}, the schemes have different optimization efforts, which are listed as a function of the circuit depth ($d=3, 4, 5$). While the \textsc{UDS} approach converges after about $4 \times 10^3$ iterations, the \textsc{CDS} simulations do not converge (n.c.) to the specified residual tolerance of $10^{-6}$. Choosing $\beta=0.5$ enables convergence and the results agree remarkably well with the analytical reference (red-line) in \Fig{fig:Results:DiffConv:Pe3}. Compared to the \textsc{UDS} method, \textsc{FB$_{0.5}$} is accompanied by an increase in the optimization steps for $\text{VQA}$ and $\text{FD}$ by $30\%$ and by a slightly higher requirement on the expressiveness of the approach, which is indicated by an increased trace distance. Based on these findings, the following two subsections are restricted to the \textsc{UDS} for the investigation of nonlinear momentum transfer studied for uni- and bidirectional inviscid cases in Secs.~\ref{sec:Results:Riemann}, \ref{sec:Results:saw}, and a case governed by the viscous Burgers' equation, cf. Sec.~\ref{sec:Results:Burgers}.

\begin{table}[htbp]
   \centering
   \caption{Deviations of \textsc{VQA} ({$\pmb{d=4}$}) and \textsc{FD} results in the $\pmb{\text{l}_2}$-norm ($\pmb{\varepsilon_{\text{l}_2}}$) and trace distance ($\pmb{\varepsilon_{\text{tr}}}$) obtained for the steady advection-diffusion problem depicted in \Fig{fig:Results:DiffConv}.}
    \begin{tabular}{@{}lcccccc@{}}
        \toprule
        Convection scheme  & \multicolumn{3}{c}{$\text{l}_2$-norm $\varepsilon_{\text{l}_2}$}  & \multicolumn{3}{c}{trace distance $\varepsilon_{\text{tr}}$} \\ 
        \midrule
        \multicolumn{1}{r}{\textit{P\'{e}clet} Number}&\multicolumn{1}{l}{$Pe_{\Delta x}=0.3$}& $Pe_{\Delta x}=3$ &$Pe_{\Delta x}=30$& \multicolumn{1}{|l}{$Pe_{\Delta x}=0.3$}& $Pe_{\Delta x}=3$ & $Pe_{\Delta x}=30$   \\
        \midrule
        UDS  $\mathcal{O}(\Delta x)$        & \multicolumn{1}{c}{$2.21 \times 10^{-5}$}& \multicolumn{1}{c}{$1.94 \times 10^{-5}$} &\multicolumn{1}{c}{$3.14\times 10^{-2}$}&\multicolumn{1}{|c}{$1.35 \times 10^{-5}$}&\multicolumn{1}{c}{$1.87 \times 10^{-5}$} &\multicolumn{1}{c}{$3.14 \times 10^{-2}$}\\
        CDS $\mathcal{O}(\Delta x^2)$  & \multicolumn{1}{c}{$2.00 \times 10^{-5}$}& \multicolumn{1}{c}{$ 1.44 \times 10^{-4}$} &\multicolumn{1}{c}{$1.34$}&\multicolumn{1}{|c}{$1.33\times 10^{-5}$}&\multicolumn{1}{c}{$2.15 \times 10^{-1}$} &\multicolumn{1}{c}{$6.94 \times 10^{-1}$}\\
         FB$_{0.5}$ $\mathcal{O}(\Delta x)$        &- &$2.57 \times 10^{-5}$ &- &\multicolumn{1}{|c}{-} &$3.33\times 10^{-1}$&-\\
        LUDS $\mathcal{O}(\Delta x^2)$  & \multicolumn{1}{c}{$3.22 \times 10^{-5}$
        }&- & \multicolumn{1}{c|}{$4.15 \times 10^{-4}$}&\multicolumn{1}{c}{$ 2.05 \times 10^{-5}$}&-& \multicolumn{1}{c}{$4.15 \times 10^{-4}$}\\
        QUICK $\mathcal{O}(\Delta x^3)$  & \multicolumn{1}{c}{$3.98 \times 10^{-5}$
        }&- & \multicolumn{1}{c|}{$1.62 \times 10^{-2}$}&\multicolumn{1}{c}{$2.53 \times 10^{-5}$}&-& \multicolumn{1}{c}{$1.49 \times 10^{-2}$}\\
        \bottomrule
    \end{tabular}
    \label{tab:Results:DIffConv:Errors}
\end{table}

\begin{table}[htbp]
   \centering
   \caption{Computational characteristics for selected \textsc{FB} convection schemes employed for the steady advection-diffusion problem ($\pmb{Pe_{\Delta x} = 3}$) depicted in \Fig{fig:Results:DiffConv}.}
   \resizebox{\textwidth}{!}{%
    \begin{tabular}{@{}lccccccccc@{}}
        \toprule
        Conv. scheme  & \multicolumn{3}{c}{UDS $(\beta=0)$}  & \multicolumn{3}{c}{FB$_{\beta}$ $(\beta=0.5)$} & \multicolumn{3}{c}{CDS $(\beta=1)$} \\ 
        \midrule
        \multicolumn{1}{r}{Circuit Depth}&\multicolumn{1}{c}{$d=3$}&\multicolumn{1}{c}{$d=4$}&$d=5$& \multicolumn{1}{|c}{$d=3$}&\multicolumn{1}{c}{$d=4$}&$d=5$ &\multicolumn{1}{|c}{$d=3$}&\multicolumn{1}{c}{$d=4$} &$d=5$  \\
        \midrule
       Opt. steps $i$& \multicolumn{1}{c}{$\frac{3969^\text{VQA}}{273^\text{FD}}$}&\multicolumn{1}{c}{$\frac{3776^\text{VQA}}{273^\text{FD}}$}&\multicolumn{1}{c}{$\frac{3051^\text{VQA}}{273^\text{FD}}$}&\multicolumn{1}{|c}{$\frac{4917^\text{VQA}}{350^\text{FD}}$}& \multicolumn{1}{c}{$\frac{6362^\text{VQA}}{350^\text{FD}}$}&\multicolumn{1}{c}{$\frac{6145^\text{VQA}}{350^\text{FD}}$}&\multicolumn{1}{|c}{n.c.
       }&  \multicolumn{1}{c}{ n.c.
       }& \multicolumn{1}{c}{n.c.
       }\\
       l$_2$-error $\varepsilon_{\text{l}_2} \,[\times 10^{-5}]$ & \multicolumn{1}{c}{$2.39$}&\multicolumn{1}{c}{$1.94$}&\multicolumn{1}{c}{$10.85$}&\multicolumn{1}{|c}{$2.45$}&$2.57$  &\multicolumn{1}{c}{$19.70$}&\multicolumn{1}{|c}{$31.91$}
       &$14.36$&\multicolumn{1}{c}{$19.15$}\\
       trace dist. $\varepsilon_{\text{tr}}$ & \multicolumn{1}{c}{$3.14\times 10^{-3}$}& \multicolumn{1}{c}{$1.87 \times 10^{-5}$}&\multicolumn{1}{c}{$3.18\times 10^{-3}$}&\multicolumn{1}{|c}{$0.33$}& $0.33$&\multicolumn{1}{c}{$0.33$}&\multicolumn{1}{|c}{$0.22$}& $0.22$&\multicolumn{1}{c}{$0.22$}\\
        \bottomrule
    \end{tabular}
}
\label{tab:OptPe3}
\end{table}

\subsection{Unidirectional Inviscid Nonlinear Convective Transport}
\label{sec:Results:Riemann}
The third application focuses on the unsteady ($a_1=0, a_2=1$), inviscid ($a_3=0$), nonlinear convection $a_4=y=v$ problem with $a_5=f=0$. In this case, the flow is unidirectional and the velocity is always positive. The left boundary is assigned to a Dirichlet inflow condition $v\big|_{x=0}=2$ and the right boundary is assigned to a Neumann outflow condition $\partial v/ \partial x \big|_{x=1}=0$. The latter is approximated by a first-order \textsc{FD} approximation, viz.  
$(v_{N_\text{p}}-v_{N_\text{p}-1})/\Delta x=0$. The employed initial velocity field features a discontinuity and reads $v(x,t=0) =2$ for $x\leq 0.75$ and $v(x,t=0) =1$ for $x>0.75$. The simulations employ the \textsc{UDS}~\eqref{eq:J_UDS} throughout $N_\text{t}+1=30$ time steps on $N_\text{p}=16$ and $N_\text{p}=64$ interior grid points for $n=4$ and $n=6$ qubits. A constant time step $\Delta t$ is applied with \textit{Courant} numbers of $v|_{x=0}\, \Delta t/\Delta x= 0.26$ ($n=4$) and $1.0$ ($n=6$). The \textsc{VQA} ansatz employs a circuit depth of $d=5$ ($n=4$) and $d=8$ ($n=6$). The convergence criterion of the optimization is set to \texttt{tol}$=10^{-7}$, limited to a maximum number of $300$ iterations. The optimization is advanced explicitly in time solving 
\begin{equation}
\min\limits_{\upb} J(\upb), \quad  \text{where } \quad J(\upb)  = a_4 \overbrace{J_\text{UDS}}^{\eqref{eq:J_UDS}}   \underbrace{-J_{\text{DN}_\text{UDS}}+J_\text{S}}_{\text{b.c. influenced \cite{Over2024}}}+ J_\text{P} \, . 
\end{equation}
Periodic boundary contributions of the \textsc{UDS} are corrected by $J_{\text{DN}_\text{UDS}}$ as discussed in Ref.~\cite{Over2024}, and non-zero Dirichlet boundary values are introduced through an additional contribution to the source term~$J_\text{S}$, whereas such term does not occur on the right end for the combination of \textsc{UDS} and positive velocities. 

\begin{figure}[htbp]
    \centering
    \begin{subfigure}[t]{0.475\textwidth}
       \includegraphics{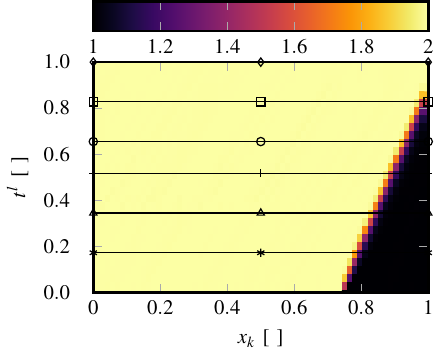}
        \caption{\textsc{VQA} predicted velocity evolutions for $\pmb{N_\text{p}=64}$ with lines and symbols marking the equidistant time instants used  to extract spatial velocity profiles in Subfig.~(\subref{fig:Results:Riemann:profiles}).}
        \label{fig:Results:Riemann:results}
    \end{subfigure}
    \hfill
    \begin{subfigure}[t]{0.475\textwidth}
        \centering
         \includegraphics{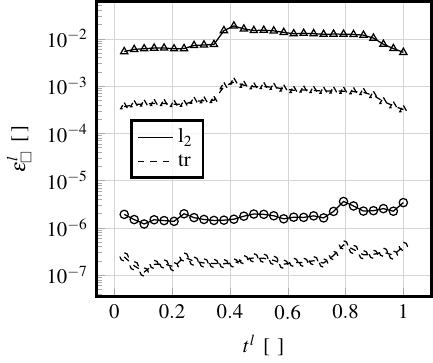}
        \caption{Temporal evolution of the error measures $\pmb{\varepsilon_{\text{l}_2}}$ and $\pmb{\varepsilon_\text{tr}}$ w.r.t. classical \textsc{FD} for $\pmb{n=4}$ ($\pmb{\circ}$) and $\pmb{n=6}$ ($\pmb{\triangle}$) qubits.}
        \label{fig:Results:Riemann:errors}
    \end{subfigure}
    \vskip\baselineskip
    \begin{subfigure}[t]{0.95\textwidth}
         \includegraphics{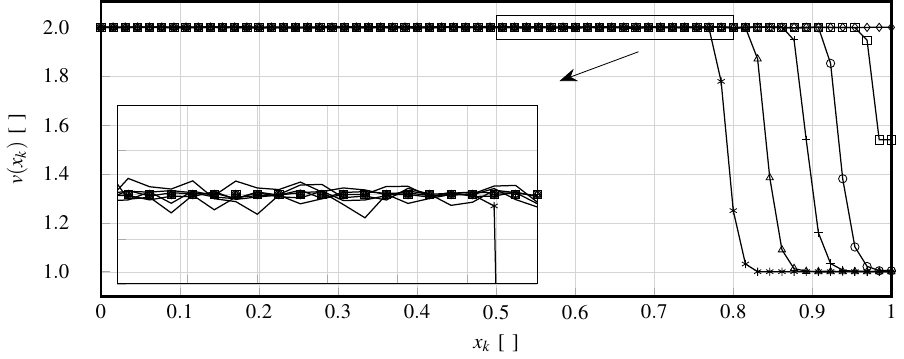}
        \caption{Spatial velocity profiles computed by the \textsc{VQA} (lines) and a \textsc{FD} method (symbols) for $\pmb{N_\text{p}=64}$ at the six time instants given in Subfig.~(\subref{fig:Results:Riemann:results}).}
        \label{fig:Results:Riemann:profiles}
    \end{subfigure}
    \caption{Results for the inviscid nonlinear convection of a shock wave obtained from the \textsc{VQA} and \textsc{FD} method  using a first-order \textsc{UDS} for the approximation of convective fluxes on a grid with $\pmb{N_\text{p}=64}$ interior points. The temporal evolution of the error measures in Subfig.~(\subref{fig:Results:Riemann:errors}) includes results for $\pmb{N_\text{p}=16}$ ($\pmb{\circ}$) and $\pmb{N_\text{p}=64}$ ($\pmb{\triangle}$) interior grid points.}
    \label{fig:Results:Riemann}
\end{figure}

Figure~\ref{fig:Results:Riemann} illustrates the results obtained from the two resolutions when using the first-order \textsc{UDS} approach. Using $n=6$ qubits, the spatio/temporal evolution of the \textsc{VQA} predicted velocity is displayed in \Fig{fig:Results:Riemann:results} where the abscissa indicates the spatial location of the $N_\text{p} = 64$ grid points and the ordinate denotes the normalized time. Figure~\ref{fig:Results:Riemann:profiles} compares the spatial velocity profiles for six exemplary time instants obtained from the \textsc{VQA} approach (lines) and the classical \textsc{FD} method (symbols) for $n=6$. Figure~\ref{fig:Results:Riemann:errors} sketches the corresponding evolution of the two error measures along a normalized time horizon (\ref{eq:Results:errors}) for $N_\text{p} = 16$ ($n = 4$) and $N_\text{p} = 64$ ($n = 6$) interior grid points. 

Analyzing the results, the use of the diffusive first-order upwind scheme reveals no instabilities across the shock and the predictive agreement between the classical and the \textsc{VQA} results again seems excellent, as indicated by \Fig{fig:Results:Riemann:profiles}. Similar to the study in Sec.~\ref{sec:results:material}, a different quality of the predictive agreement obtained for coarse and the fine resolution is outlined by the l$_2$ error ($\varepsilon_{\text{l}_2}$) and trace distance error ($\varepsilon_\text{tr}$) depicted in \Fig{fig:Results:Riemann:errors}. The l$_2$ errors maintain levels of the order of $\varepsilon_{\text{l}_2}^l \approx 10^{-6}$ ($n=4$) and $\varepsilon_{\text{l}_2}^l \approx 10^{-2}$ ($n=6$) with temporal averages of $\bar\varepsilon_{\text{l}_2}=1.92 \times 10^{-6}$ ($n=4$) and $\bar\varepsilon_{\text{l}_2}=1.04 \times 10^{-2}$ ($n=6$). The trace distance error is of one order of magnitude lower with temporal averages of $\bar\varepsilon_\text{tr}=2.25 \times 10^{-7}$ and $\bar\varepsilon_\text{tr}=6.68 \times 10^{-4}$ for $n=4$ and $n=6$ qubits, respectively. The curves for $\varepsilon_{\text{l}_2}$ and $\varepsilon_\text{tr}$ in \Fig{fig:Results:Riemann:errors} again increase with the number of qubits~$n$, but remain approximately constant over time. The different levels agree with the findings of Sec.~\ref{sec:results:material} and are attributed to limitations of the ansatz and resulting challenges in the convergence of the optimization process. As detailed in \Fig{fig:Results:Riemann:profiles}, the problem is (again) most pronounced in zones with weakly varying solution values. However, the overall solution quality is still satisfactory, and the corresponding error levels for $n=6$ are still slightly below the values reported in the literature for five qubits \cite{Sato2021, Leong2022}. The deviations are oscillatory in nature and indicate non-local correlations. These occur particularly for larger depths~$d$ and might be addressed by suitable regularization techniques. In addition, the use of classical shadows \cite{Sack22}, exploiting classical neural networks \cite{Friedrich22}, adding entangled ancilla qubits \cite{Yao25}, or ansatzes targeting certain tensor network quantum states \cite{Barthel25} are expected to improve the convergence of the optimization and increase the accuracy of the obtained results.

\subsection{Bidirectional Inviscid Nonlinear Convective Transport}
\label{sec:Results:saw}
The fourth case scrutinizes the presented method for inhomogeneous flow directions, which represents an important algorithmic detail of the proposed approach. 
It features an unsteady ($a_1=0$, $a_2=1$), inviscid ($a_3=0$), nonlinear, bidirectional convection ($a_4=y=v$) problem with $a_5=f=0$. For this purpose, the boundaries are assigned to periodic conditions, and the initial velocity field features an opposing flow of non-symmetric triangular shape with a maximum magnitude of $v_\text{max}=1$, viz. 
\begin{equation}
    v(x,t=0) = 
    \begin{cases}
        \;\; 4x         \ \ \ \qquad \quad x<0.25L \\
        -4x+2         \ \ \ \ \quad x\geq 0.25L \land x<0.5L \\
        -2x+1           \quad \quad x\geq 0.5L \land x<0.75L \\
        \;\; 2x-2     \quad \quad x \geq 0.75L \land x<L \, .
    \end{cases}
\end{equation}
The simulations cover $N_\text{t}+1=30$ ($n=4$) and $N_\text{t}+1=16$ ($n=6$) time steps with a constant non-dimensional time step $\Delta t v_\text{max}/\Delta x =0.075$ ($n=4$) and $\Delta t v_\text{max}/\Delta x =0.03$ ($n=6$), respectively. The discretization employs again the \textsc{UDS}~\eqref{eq:UDS} based on a mask function implementation for the approximation of convective kinematics, which inherently uses both backward and forward differences. The \textsc{VQA} simulations are conducted with an ansatz depth of $d=6$ ($n=4$) and $d=8$ ($n=6$), an optimization convergence criteria of $\texttt{tol}=10^{-7}$, and a maximum number of iterations of $300$. In this regard, the solution at each time step follows from 
\begin{equation}
\min\limits_{\upb} J(\upb), \quad \text{where } \quad 
J(\upb)  =-a_3 \overbrace{J_\text{cent-2}}^{\eqref{eq:seconddiff_quantum}}+ a_4 \overbrace{J_\text{UDS}}^{\eqref{eq:J_UDS}}  + J_\text{P} + J_\text{S}\, .
\label{eq:cost_bidirec}
\end{equation}
Because the circuits of Sec.~\ref{sec:Quantum} comply with the periodic setting, no boundary corrections of the cost function~\eqref{eq:cost_bidirec} are required, and the source term~$J_\text{S}$ only accounts for contributions related to the temporal discretization.

Figure~\ref{fig:Results:Saw} collects the results structured as in the previous figures, i.e., \Fig{fig:Results:Saw:results} depicts the spatio/temporal transport, \Fig{fig:Results:Saw:profiles} shows velocity profiles at selected time instances, and \Fig{fig:Results:Saw:errors} outlines the evolution of error measures~\eqref{eq:Results:errors} along a normalized time horizon obtained for $n=4$~($\circ$) and $n=6$~($\triangle$) qubits. The adequate modeling of the convective kinematics by the presented masking approach is confirmed by the figures. The faster transport associated with the larger velocity magnitude on the left part of the domain is clearly visible for the \textsc{VQA} and the \textsc{FD} solutions, and results  seem to display a very good predictive agreement. Moreover, the convective transport is characterized by an increase of spatial gradients at the two downstream fronts and a reduction of gradients along the upstream fronts, as shown in Fig.~\ref{fig:Results:Saw:profiles}. The predictive agreement varies only moderately over time, as the profiles in \Fig{fig:Results:Saw:profiles} and the error measures in \Fig{fig:Results:Saw:errors} show. Mind that the error levels basically agree with the findings of previous cases and are again augmented for the fine grid, due to difficulties in converging the optimizer in combination with the employed ansatz function.
\begin{figure}[htbp]
    \centering
    \begin{subfigure}[t]{0.475\textwidth}
       \include{Res-Burg-Saw}
        \caption{\textsc{VQA} predicted velocity contours for $\pmb{N_\text{p}=64}$ with lines and symbols marking the equidistant time instants used to extract spatial velocity profiles in Subfig.~(\subref{fig:Results:Saw:profiles}).}
        \label{fig:Results:Saw:results}
    \end{subfigure}
    \hfill
    \begin{subfigure}[t]{0.475\textwidth}
        \centering
        \tikzsetnextfilename{Error-burg-gauss}
\begin{tikzpicture}
\begin{axis}[
tick pos=left,
xmajorgrids,
ymajorgrids,
%ymin=1.0e-3, ymax=1.0e-1,
log basis y={10},
legend style={draw=black!15!black,legend cell align=left, fill opacity=0.8, draw opacity=1, text opacity=1, at={(0.1,0.5)}, anchor=west},
ymode=log,
%xmode=log,
xlabel={$t^l$ [\   ]},
ylabel={$\varepsilon_{\square}^l$ [\   ]},
width=0.75\textwidth,  
height=5cm, 
]

\addlegendimage{semithick, black}
\addlegendentry{$\text{l}_2$}

\addlegendimage{semithick,dashed, black}
\addlegendentry{\text{tr}}

%%%%%%%%%%% n=6
\addplot [semithick, mark=triangle]
table[col sep = comma, row sep=crcr] {%
0.066666667	,   0.010626393506535528 \\
0.133333333	,	0.016621773224904657 \\
0.2	,   0.01955800642485527  \\
0.266666667	,   0.016892254153121896 \\
0.333333333	,   0.015308901636427877 \\
0.4	,   0.014008352349391533 \\
0.466666667	,   0.014290838738347493 \\
0.533333333	,   0.015860964443039936 \\
0.6	,   0.018918618532479257 \\
0.666666667	,   0.019856694742230844 \\
0.733333333	,   0.02085185680398557  \\
0.8	,   0.021655016658986444 \\
0.866666667	,   0.021651899073770128\\
0.933333333	,   0.02097107801175638\\
1	, 0.0216550166589864\\
};
%\addlegendentry{l$_2$}
%%%%%%%%%%% n=6
\addplot [semithick, mark=triangle, dashed]
table[col sep = comma, row sep=crcr] {%
0.066666667	,   0.002914792162070801  \\
0.133333333	,   0.004570446107752222  \\
0.2	,   0.005390664332628502  \\
0.266666667	,   0.00466689948434678   \\
0.333333333	,   0.004239357920039001  \\
0.4	,   0.0038882580940712073  \\
0.466666667	,   0.003969490168062962   \\
0.533333333	,   0.004402268435023048   \\
0.6	,   0.005252436884385285   \\
0.666666667	,   0.005534381746002658  \\
0.733333333	,   0.005832680208619877  \\
0.8	 ,  0.006077197877595322  \\
0.866666667	,   0.006094253968594293 \\
0.933333333	,   0.00591889165159141  \\
1	, 0.0054979446697327 \\
};
%\addlegendentry{\text{tr}}

%%%%%%%%%%% n=4
\addplot [semithick, mark=o]
table[col sep = comma, row sep=crcr] {%
0.03	, 2.3863726298776863e-07	\\
0.07	, 2.6496504791837949e-07	\\
0.10	, 2.9202278182882552e-07	\\
0.14	, 3.6640038648037274e-07	\\
0.17	, 3.2449824799807287e-07	\\
0.21	, 3.4801538575015208e-07	\\
0.24	, 3.4863212112985823e-07	\\
0.28	, 4.0874437957529990e-07	\\
0.31	, 3.7137766920976146e-07	\\
0.34	, 3.7723432954105064e-07	\\
0.38	, 3.9732156210285163e-07	\\
0.41	, 5.0873261007274914e-07	\\
0.45	, 4.3341201805728778e-07	\\
0.48	, 4.2425143399002043e-07	\\
0.52	, 5.7258906702393924e-07	\\
0.55	, 6.6334008109790122e-07    \\
0.59	, 7.2863158762714491e-07	\\
0.62	, 7.1894973595930145e-07	\\
0.66	, 7.0380748614048474e-07	\\
0.69	, 6.1152897115142720e-07	\\
0.72	, 5.6864213099411565e-07	\\
0.76	, 5.5896675297435730e-07	\\
0.79	, 4.9160562157923272e-07	\\
0.83	, 5.7285563865394773e-07	\\
0.86	, 6.0745749978068115e-07	\\
0.90	, 6.0113557239437655e-07    \\
0.93	, 5.9938624566312421e-07    \\
0.97	, 6.2760757667525656e-07	\\
1	    , 7.2356535381777160e-07 \\
};
%\addlegendentry{l$_2$}
%%%%%%%%%%% n=4
\addplot [semithick, mark=o, dashed]
table[col sep = comma, row sep=crcr] {%
0.03	,	1.2467205919833117e-07	\\
0.07	,	1.4136482746161726e-07  \\
0.10	,	1.5769906706002895e-07	\\
0.14	,	2.0212918469683135e-07	\\
0.17	,	1.8005141576520461e-07	\\
0.21	,	1.9256544088304171e-07	\\
0.24	,	1.9140888122715249e-07	\\
0.28	,	2.2696772533249583e-07	\\
0.31	,	2.1020674675205178e-07	\\
0.34	,	2.1178529267835461e-07	\\
0.38	,	2.2500259981366246e-07	\\
0.41	,	2.9162147329421230e-07	\\
0.45	,	2.4845200777848162e-07	\\
0.48	,	2.4439721505130420e-07	\\
0.52	,	3.1327897372288464e-07	\\
0.55	,	3.7223088731756200e-07	\\
0.59	,	4.1829552792364952e-07	\\
0.62	,	4.1696633635858362e-07	\\
0.66	,	4.0590281700719105e-07	\\
0.69	,	3.6102656840043418e-07	\\
0.72	,	3.3750389972049370e-07	\\
0.76	,	3.3219898953449659e-07	\\
0.79	,	2.9276136613471060e-07	\\
0.83	,	3.4369714840265502e-07	\\
0.86	,	3.6194794778499361e-07	\\
0.90	,	3.5199558171682703e-07  \\
0.93	,	3.5544813996185551e-07	\\
0.97	,   3.8194673114944388e-07  \\
1.00	,	4.4128575608017586e-07  \\
};
\end{axis}
\end{tikzpicture}
        \caption{Temporal evolution of the error measures $\pmb{\varepsilon_{\text{l}_2}}$ and $\pmb{\varepsilon_\text{tr}}$ w.r.t. classical \textsc{FD} for $\pmb{n=4}$ ($\pmb{\circ}$) and $\pmb{n=6}$ ($\pmb{\triangle}$) qubits.}
        \label{fig:Results:Saw:errors}
    \end{subfigure}
    \vskip\baselineskip
    \begin{subfigure}[t]{0.95\textwidth}
    \include{Profile-Burg-Saw}
    \caption{Spatial velocity profiles computed by the \textsc{VQA} (lines) and the \textsc{FD} method (symbols) for $\pmb{N_\text{p}=64}$ at the four time instants marked in Subfig.~(\subref{fig:Results:Saw:results}).}
    \label{fig:Results:Saw:profiles}
    \end{subfigure}
    \caption{Results for the inviscid, nonlinear,  bidirectional convection obtained from the \textsc{VQA} and \textsc{FD} method using first-order \textsc{UDS} to approximate convective fluxes on a grid with $\pmb{N_\text{p}=64}$ interior points. The temporal evolution of the error measures in Subfig.~(\subref{fig:Results:Saw:errors}) includes results for $\pmb{N_\text{p}=16}$ ($\pmb{\circ}$) and $\pmb{N_\text{p}=64}$ ($\pmb{\triangle}$) interior grid points.}
    \label{fig:Results:Saw}
\end{figure}

\subsection{Viscous Nonlinear Convective Transport (Burgers' Equation)}
\label{sec:Results:Burgers}
The fifth test case adds viscous forces to the previous nonlinear momentum ($a_4 = y = v$) transport ($a_1=a_5=f=0, \, a_2=1$). To this end, the \textit{Reynolds} number is set to $Re=100$ ($a_3 = 0.01$). The investigated example is subjected to periodic boundary conditions $v\big|_{x=0} = v\big|_{x=1}$ and initialized with a Gaussian distribution in the second quarter of the spatial domain, viz. $v(x,t=0)=v_\text{max} \cdot \exp\big({-(10\,x-3.5)^4}\big)$ with $v_\text{max}=1$. The spatial discretization employs $N_\text{p}=16$ and $N_\text{p}=64$ interior grid points or $n=4$ and $n=6$ qubits, respectively. The simulation evolves over $N_\text{t}+1=25$ time steps with \textit{Courant} numbers (${v_\text{max} \, \Delta t}/{\Delta x}$) of $0.28$ and $1$, resulting in $\Delta t =0.0163$ and $\Delta t =0.0154$ for $n=4$ and $n=6$, respectively. Each time instant requires solving the optimization problem 
\begin{equation}
\min\limits_{\upb} J(\upb), \quad \text{where } \quad 
J(\upb)  =-a_3 \overbrace{J_\text{cent-2}}^{\eqref{eq:seconddiff_quantum}}+ a_4 \overbrace{J_\text{UDS}}^{\eqref{eq:J_UDS}}  + J_\text{P} + J_\text{S}\, .
\end{equation}
Due to the periodic setting, no corrections are needed for the boundary contributions to the objective functional. Diffusive and transient terms are treated implicitly, such that the contributions $J_\text{P}$ and $J_\text{S}$ only account for parts of the temporal discretization and the convective term as described in Sec.~\ref{sec:Quantum}. The \textsc{VQA} ansatz employs a circuit depth of either ${d=5}$ ($n=4$) or ${d=8}$ ($n=6$) and the convergence criterion of the optimization is \texttt{tol}$=10^{-7}$, limited to a maximum iteration number of $300$.

\begin{figure}[htbp]
    \centering
    \begin{subfigure}[t]{0.475\textwidth}
        \includegraphics{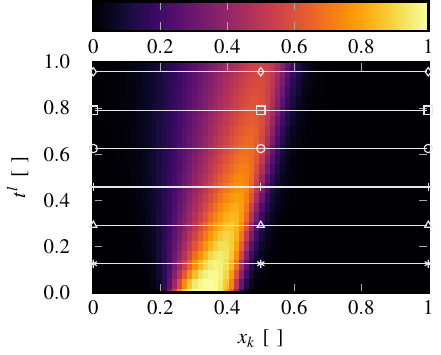}
        \caption{\textsc{VQA} simulations for $\pmb{N_p=64}$ with lines and symbols marking the equidistant time instants used to extract spatial velocity profiles in Subfig.~(\subref{fig:Results:Burgers:profiles}).}
        \label{fig:Results:Burgers:results}
    \end{subfigure}
        \hfill
    \begin{subfigure}[t]{0.475\textwidth}
        \centering
        \includegraphics{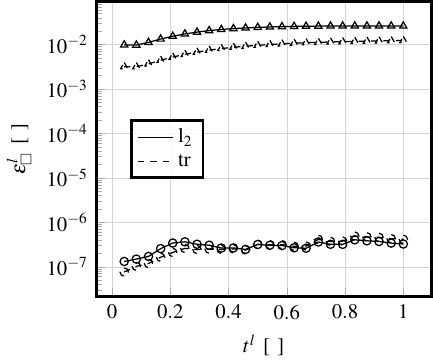}  
        \caption{Temporal evolution of the error measures $\pmb{\varepsilon_{\text{l}_2}}$ and $\pmb{\varepsilon_\text{tr}}$ w.r.t. classical \textsc{FD} for $\pmb{n=4}$ ($\pmb{\circ}$) and $\pmb{n=6}$ ($\pmb{\triangle}$) qubits.}
        \label{fig:Results:Burgers:errors}
    \end{subfigure}
    \vskip\baselineskip
    \begin{subfigure}[t]{0.95\textwidth}
        \includegraphics{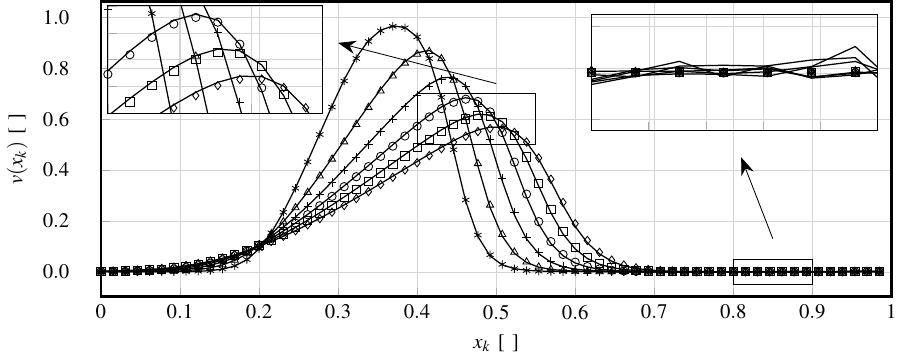}
        \caption{Velocity profiles computed by the \textsc{VQA} (lines) and \textsc{FD} method (symbols) for $\pmb{N_\text{p}=64}$ at the six time instants indicated in Subfig.~(\subref{fig:Results:Burgers:results}).}
        \label{fig:Results:Burgers:profiles}
    \end{subfigure}
    \caption{Temporal evolution of the \textsc{VQA} results for the nonlinear %transport problem governed by the 
    Burgers' equation at $\pmb{Re=100}$ subject to periodic boundary conditions and an initial Gaussian distribution $\pmb{v(x,t=0) =exp\big(-(10\,x-3.5)^4\big)}$. The temporal evolution of the error measures in Subfig.~(\subref{fig:Results:Burgers:errors}) include results for $\pmb{N_\text{p}=16}$ and $\pmb{N_\text{p}=64}$ interior grid points.}
    \label{fig:Results:Burgers}
\end{figure}

The comparison between the \textsc{VQA} and the \textsc{FD} results is displayed in \Fig{fig:Results:Burgers}, which is decomposed into the same arrangement of subfigures as of the previous subsection, i.e., space/time \textsc{VQA} results, temporal evolution of the error measures, and velocity profiles comparisons. Mind that all temporal axes refer to a normalized time. In line with the previous case, the nonlinear transport of the initially inhomogeneous (Gaussian) velocity distribution reveals a stretching of the upstream front and an increase of the spatial gradient at the downstream front, cf.~\Fig{fig:Results:Burgers:results}. The employed \textsc{UDS} approximation additionally broadens both fronts by introducing numerical diffusion, as explained in Sec.~\ref{sec:results:steady_diff_conv}. Nonetheless, the evolution of the initially symmetric distribution into a non-symmetric triangular distribution indicates a successful representation of the nonlinear convective kinematics, which is also confirmed by the velocity profiles in \Fig{fig:Results:Burgers:profiles}. 
The results of the \textsc{VQA} and the \text{FD} methods visually agree very well in \Fig{fig:Results:Burgers:profiles}. However, although both error measures~\eqref{eq:Results:errors} depicted in \Fig{fig:Results:Burgers:errors} converge with successive time steps, the time averaged error norms fall below $10^{-6}$ for $n=4$ but persists at approximately $10^{-2}-10^{-3}$ for $n=6$. Again the observed increase of error measures is attributed to convergence difficulties of the optimizer in connection with the chosen ansatz in regions where the flow is virtually constant.

\subsection{Linear Wave Equation}
\label{sec:results:wave}
The final application addresses the transport of pressure perturbations~$y=p$ governed by the hyperbolic linear wave equation $\big(a_1=Ma^2$, $a_3=1$, and $a_2=a_4=a_5=f=0\big)$. For this application the \textit{Mach} number is set to $Ma^2=1$, both boundaries are modeled as reflective walls by the Dirichlet conditions $p\big|_{x=0}=p\big|_{x=1}=0$, and the initial pressure distribution is set to the Gaussian $p(x,t=0) = \exp\big({-(10\,x-3.5)^4}\big)$. The time-implicit simulations employ a fixed time step of $\Delta t = 0.05$ evolving in $N_\text{t}+1=30$ time steps on grids with $N_\text{p}=16$ ($n=4$) and $N_\text{p}=64$ ($n=6$) interior grid points. The pressure is encoded using the bricklayer ansatz of Sec.~\ref{sec:hybrid} with depths $d=4$ and $d=7$ for $n=4$ and $n=6$, respectively. The convergence criterion for the optimizations is \texttt{tol}$=10^{-7}$, limited to a maximum number of $300$ iterations. 

To enforce the homogeneous Dirichlet boundary conditions, corrections of the periodic contribution from the central differences $J_\text{cent-2}$ are required~\cite{Over2024} and marked by $J_{\text{DN}_\text{cent-2}}$. Analogous to the previous cases, the second derivative w.r.t. time, cf.~\Eq{eq:timeDiscr}, is attributed to a potential term~$J_\text{P}$ and source term~$J_\text{S}$. Apart from the latter contribution to the linear form~$F(y)$, the spatial differencing schemes are symmetric and can be realized fully implicit on the normalized time interval $[0,1]$. The time evolution follows from the solution of an optimization problem for each time instant $l$, viz. 
\begin{equation}
\min\limits_{\upb} J(\upb), \quad  \text{where } \quad J(\upb)  = -a_3 \overbrace{J_\text{cent-2}}^{\eqref{eq:seconddiff_quantum}} + a_3\underbrace{J_{\text{DN}_\text{cent-2}}}_{\text{b.c. \cite{Over2024}}} + J_\text{P} + J_\text{S}\,.
\end{equation}

\begin{figure}[htbp]
    \centering
    \begin{subfigure}[t]{0.475\textwidth}
        \includegraphics{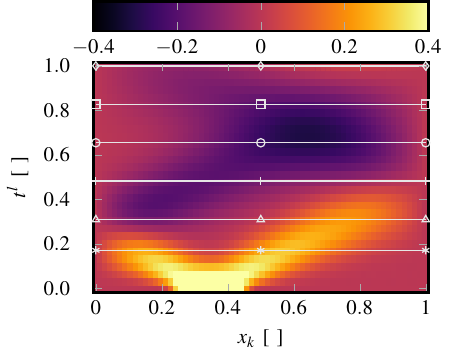}
        \caption{\textsc{VQA} simulation on $\pmb{N_\text{p}=64}$ interior points with lines and symbols marking the equidistant time instants used to extract spatial pressure profiles in Subfig.~(\subref{fig:Results:Wave:profils}).}
        \label{fig:Results:Wave:results}
    \end{subfigure}
        \hfill
     \begin{subfigure}[t]{0.475\textwidth}
        \centering
        \includegraphics{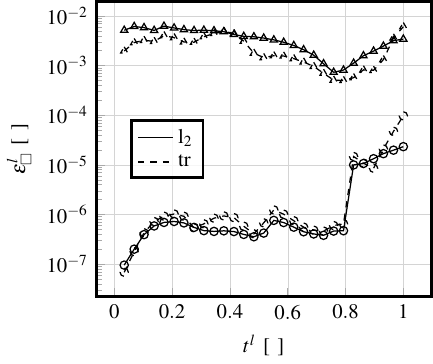}    
        \caption{Temporal evolution of the error measures $\pmb{\epsilon_{\text{l}_2}}$ and $\pmb{\epsilon_\text{tr}}$ w.r.t. classical \textsc{FD} for $\pmb{n=4}$ ($\pmb{\circ}$) and $\pmb{n=6}$ ($\pmb{\triangle}$) qubits.}
        \label{fig:Results:Wave:errors}
    \end{subfigure}
    \vskip\baselineskip
    \begin{subfigure}[t]{0.95\textwidth}
        \includegraphics{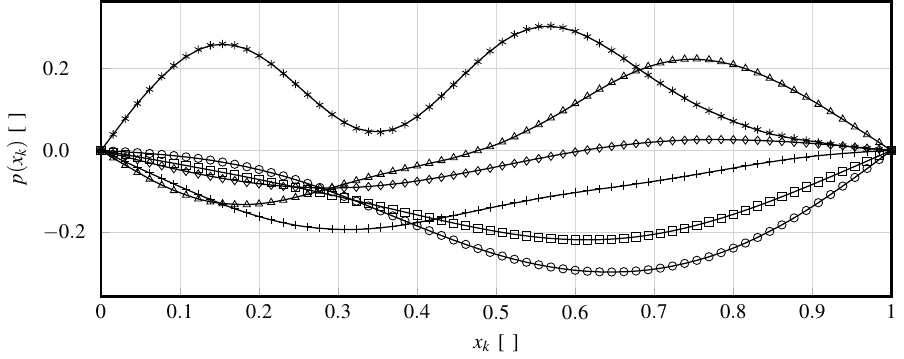}
        \caption{Pressure distributions recovered by the \textsc{VQA} (lines) and \textsc{FD}-reference (symbols) for $\pmb{N_\text{p}=64}$ at the six time instants indicated in Subfig.~(\subref{fig:Results:Wave:results}).}
        \label{fig:Results:Wave:profils}
    \end{subfigure}
     \caption{Temporal evolution of the \textsc{VQA} results for the linear transport problem governed by the wave equation subjected to homogeneous Dirichlet boundary conditions and a Gaussian initial pressure. The temporal evolution of the error measures in Subfig.~(\subref{fig:Results:Wave:errors}) include results for $\pmb{N_\text{p}=16}$ ($n=4$) and $\pmb{N_\text{p}=64}$ ($n=6$) simulations.}
    \label{fig:Results:Wave}
\end{figure}

Figure~\ref{fig:Results:Wave} illustrates the results obtained for the \textsc{VQA} and the classical \textsc{FD} approach in the previously introduced manner. The spatio/temporal evolution of the \textsc{VQA}-predicted pressure is displayed in \Fig{fig:Results:Wave:results} for $N_\text{p}=64$ ($n=6$). Again, the abscissa refers to the spatial location and the ordinate to the time. The figure also indicates the six time instants that are used to compare spatial velocity profiles obtained by the \textsc{VQA} and the classical method in \Fig{fig:Results:Wave:profils}. 

The initial Gaussian distribution is clearly observed from the light color near the ordinate in \Fig{fig:Results:Wave:results}. As time evolves, the wave propagates in both directions and changes phase when being reflected at the boundaries. The reflected waves subsequently propagate in the opposite direction and interact within the third spatial quarter, resulting in a situation that is spatially antisymmetric to the initial condition. As seen from \Fig{fig:Results:Wave:profils}, the \textsc{VQA} results agree with their \textsc{FD} counterparts (symbols). Figure~\ref{fig:Results:Wave:errors} provides quantitative information about the temporal evolution of the error measures  \eqref{eq:Results:errors} for $N_\text{p}=16$ and $N_\text{p}=64$  grid points. Similarly to the previous subsections, the figure reveals an increase in the error for the finer grid. The time averaged errors are $\bar\varepsilon_{\text{l}_2} = 3.49 \times 10^{-6}$ and $\bar\varepsilon_\text{tr} = 7.15 \times 10^{-6}$ for the four qubit case and $\bar\varepsilon_{\text{l}_2}=0.034$ and $\bar\varepsilon_{\text{tr}}=0.025$ for the six qubit case. Moreover, the figure reveals an error minimum close to a level of $10^{-6}$ ($n=4$) [$10^{-3}$ ($n=6$)] when the pressure waves interact around $t \approx 0.75$.

%###########################################
%       Conclusion
%###########################################
\section{Conclusion}
\label{sec:Conclusion}
The work presents a flexible \textsc{VQA} for simulating linear and nonlinear fluid dynamic transport phenomena using a hybrid classical-quantum computing framework. The method combines \textsc{FD} approximations for differential operators with an optimization problem that minimizes the weak residual formulation of the associated governing \textsc{PDE}. To this end, the parameters of a modular ansatz circuit are trained on the classical hardware while the expensive evaluations of the objective functional are executed on the \textsc{QC}. 

The approach implicitly treats symmetric discretization contributions, while non-symmetric/biased discretizations, e.g., to approximate convective kinematics, are explicitly treated. It supports non-homogeneous diffusivity and the encoding of classical, upwind-biased schemes of first (\textsc{UDS}) and higher-order (\textsc{LUDS}, \textsc{QUICK}) into the quantum concept, which, in contrast to central differences, do not generate dispersion waves even at high \textit{Reynolds}/\textit{P\'{e}clet} numbers. The \textsc{VQA} results generally seem to be in very good agreement with those obtained using classical methods. The remarkable agreement covers a wide range of benchmark cases and transport phenomena and include different numerical approaches, for example ansatz depths, discretization schemes and boundary conditions. This demonstrates that the proposed VQA method can solve various types of transport PDEs. The agreement is particularly excellent for four qubits, while some deviations were observed for six qubits, consistent with the existing literature \cite{Sato2021,Leong2022,Over2024}. The deviations are attributed to the ansatz functions and could be addressed using an appropriate regularization of the objective functional and other strategies that help to improve the convergence of the optimization \cite{Sack22,Friedrich22,Yao25,Barthel25}. Moreover, it is demonstrated that a physically appropriate modeling, e.g., by using upwind-biased schemes for convective kinematics, reduces the optimization costs of the hybrid \textsc{VQA} approach. In particular, the use of upwind differencing methods achieves the lowest cost function contributions, the lowest ansatz depth, and the lowest optimization effort, which can be combined with a higher spatial resolution to solve accuracy problems.

The implementation encodes all non-zero diagonals of the discrete operators into individual \textsc{QNPU}s and embeds them in Hadamard test-based circuits. The resulting scalar-valued multi-objective optimal control problem is constructed with only one ancilla qubit measurement per circuit and it features remarkably low circuit counts for symmetric and Toeplitz matrices which generally appear in \textsc{CFD} applications. In addition, an optimal scaling in gate complexity w.r.t. the number of qubits is achieved, which is \textit{polylog}, due to the efficient combination of adder and potential circuits.

Future work will refer to non-uniform grids, along the route outlined for the non-constant material example, and an extension to more than one spatial dimension by tensor products of the spatial discretization. Open challenges for amplitude phase estimation in terms of $\min$/$\max$~operations and strategies to regularize the ansatz to suppress long-range correlations will also be addressed. 

%###########################################
%       Acknowledgements
%###########################################
\section*{Acknowledgments}
This publication and the current work have received funding from the \href{https://doi.org/10.3030/101080085}{European Union's Horizon Europe research and innovation program (HORIZON-CL4-2021-DIGITAL-EMERGING-02-10) under grant agreement No. 101080085 \textsc{QCFD}}. 

%###########################################
%       Author Declarations
%###########################################
\section*{Author Contributions}
\addcontentsline{toc}{section}{Author Contributions}
\textbf{Sergio~Bengoechea:} Conceptualization, Methodology, Software, Validation, Formal analysis, Investigation, Writing - original draft, Writing - review \& editing, Visualization. 
\textbf{Paul~Over:} Conceptualization, Methodology, Software, Validation, Formal analysis, Investigation, Writing - original draft, Writing - review \& editing, Visualization. 
\textbf{Dieter~Jaksch:} Project administration, Funding acquisition, Supervision, Resources, Writing - review \& editing. 
\textbf{Thomas~Rung:} Project administration, Funding acquisition, Supervision, Conceptualization, Methodology, Resources, Writing - original draft, Writing - review \& editing. 

%###########################################
%       Data Availability
%###########################################
\section*{Data availability}
\addcontentsline{toc}{section}{Data availability}
The data is available via \href{https://doi.org/10.25592/uhhfdm.16634}{https://doi.org/10.25592/uhhfdm.16634}.

%###########################################
%      BIBLIOGRAPHY
%###########################################
\bibliography{lib}

\end{document}